\documentclass[onecolumn,aps,prd,showpacs,floatfix,eqsecnum,nofootinbib]
{revtex4}
\usepackage{amsmath,graphicx}

\begin{document}
\title{Spin-Orbit Resonance and the Evolution of Compact Binary Systems} 
\author{Jeremy D. Schnittman}
\affiliation{Department of Physics, MIT, 77 Massachusetts Ave.,
Cambridge, MA 02139}

\begin{abstract}
Starting with a post-Newtonian description of compact binary
systems, we derive a set of equations that describes the evolution
of the orbital angular momentum and both spin vectors during inspiral. We find
regions of phase space that exhibit resonance behavior, characterized
by small librations of the spin vectors around a fixed
orientation. Due to the loss of energy and orbital angular
momentum through radiation reaction, systems can eventually be
captured into these resonance orientations. By investigating the
long-term evolution of compact binaries
with a variety of initial conditions, we find that the distribution in
parameter space can be strongly affected by resonance captures. This
has the effect of significantly reducing the size of search space
for gravitational wave sources, in turn improving the chances of
detecting such sources through methods of template
matching. Furthermore, by calculating the expected spin distribution
at the end of the inspiral phase, we can predict what are the most
likely initial conditions for the plunge phase, a result of great
interest for numerical relativity calculations.
\end{abstract}
\maketitle

\section{Introduction}
The inspiral and subsequent coalescence of two black holes (BH) or
neutron stars (NS) promises
to be a strong source of gravitational waves for a number of new
interferometric detectors currently being
developed\footnote{http://www.ligo.caltech.edu/ \newline
http://www.virgo.infn.it \newline
http://www.geo600.uni-hannover.de \newline
http://tamago.mtk.nao.ac.jp \newline
http://lisa.nasa.gov}. In order to
successfully detect and then analyze such sources,
we need to have an accurate description of the gravitational waveforms
they will produce. For longer signals lasting many orbital cycles
(typically NSs in the LIGO frequency band and BHs in the LISA
band), the detection methods rely heavily on a technique called
matched filtering. This method is based on the premise that we can
calculate theoretical templates of gravitational waves which in turn
are cross-correlated with the observed data in an attempt to match a small
amplitude signal buried under high background noise. Since the
anticipated signal can be hundreds or even thousands of cycles long,
it is critical that the form of the theoretical template is very
accurate or the detection could be missed. Once detected, the observed
waveform must then be compared to a larger collection of model templates in
order to fit the binary parameters and determine their statistical
confidences.

Approximate templates that do not include spin effects have been shown
to have a poor chance of matching gravitational waveforms from
spinning binaries \citep{apost95,grand03a,grand03b,buona03,grand04,pan04}.
Even if we were able to calculate the theoretical waveforms with
perfect physical accuracy, the binary black hole system is so
complicated that we would need a very large template library to give a
reasonable chance at detection \citep{buona03,buona04}. For two
spinning black holes, the parameter space is
characterized by at least 11 intrinsic variables [the masses (2), the angular
momentum vector (3), and two spin vectors (6)]. \citet{apost94} showed
that the two-spin system can be reduced in the limits of equal mass
when neglecting spin-spin interactions and later \citet{apost96} included
these terms for the equal mass, equal spin case. Recent work by Buonanno et
al.\ \citep{buona04} showed that the size of this parameter space can
also be reduced
by considering a set of \textit{quasi-physical} templates that mimic the
physical behavior of two spins with a single effective spin. \citet{grand04}
have suggested a different method of using ``spiked'' templates to expand the
search templates and simulate spin effects. While these fitting methods
greatly aid the searches for gravitational waves, they could also benefit
from additional astrophysical information about the systems that are
producing the waves, as well as their evolution up to the point where they
enter the frequency regime of the detector. The added advantage of
using strictly physical templates is the direct manner in
which they allow us to determine the intrinsic parameters of the
compact binary.

There are a number of stellar evolution models that describe the
formation of binary black hole systems, including estimates for
initial spins and kick velocities, which in turn can give the
orientation of the orbit
\citep{kalog00,belcz02,pfahl02,nutzm04}. However, there is
still a fair amount of uncertainty in the appropriate initial values
to use for inspiraling stellar mass black hole binaries. 
The mechanisms governing supermassive or
intermediate mass black hole mergers are even less
certain. In both cases, we have little or no idea what to expect the
system might look like as it enters the final stages of evolution
towards inspiral and merger. 

In addition to the compact objects formed through binary stellar
evolution, another important source of gravitational waves may be capture
binaries \citep{porte00}. These include binary systems that form in
the cores of dense
globular clusters, stellar mass and supermassive black holes in the
centers of galaxies, and supermassive black hole binaries formed
through hierarchical galactic mergers. Many of these systems should
have no a priori preference for any particular spin orientation.
It is also possible that some of these systems have already been
seen, but not through the detection of gravitational waves. A recently
discovered binary pulsar system may give important information about
the initial 
spin-orbit orientations and early evolution of inspiraling neutron
stars \citep{lyne04,wille04}. \citet{merri02} suggest that the
morphology of radio jets from active galaxies may point to recent
supermassive BH mergers and resulting changes in spin orientation. In
fact, the longer term merger history of a supermassive black hole might
also be inferred by its current mass and spin \citep{hughe03}. If we
could give a realistic prediction
of the orientation of both spins and the angular momentum vector 
relative to each other at the time of inspiral, an important link
between these different branches of astrophysics could be established. A
successful identification with an electromagnetic source would also be
critical in the confirmation of any gravitational wave detection,
especially in the early years of gravitational wave astronomy.

To further investigate these issues, we have developed a code to integrate the
post-Newtonian equations of motion and spin precession for two
spinning point masses including radiation reaction. The full evolution
of the system from its formation until merger covers an enormous range
of time scales, from hundreds of megayears to fractions of
milliseconds. To accommodate this range, we first derive an
orbit-averaged system of evolution equations that model the
orientation of the spins and angular momentum without actually following the
orbital phase of the binary. These orbit-averaged equations of motion
agree quite well with the full 2.5-order post-Newtonian equations
for modeling the evolution of the orbital angular
momentum and both spin vectors \citep{kidde95}.

With this orbit-averaged formulation, a collection of equilibrium solutions
are found in which the relative orientation of the spin vectors and
orbital angular momentum
remains fixed in time. Furthermore, we find that the majority of these
equilibrium solutions are stable, so systems nearby in parameter space
librate around the stable orbits like a spin-orbit or spin-spin
resonance. With the inclusion of radiation reaction, spin-locked
systems remain locked, following a trajectory along equilibrium
solutions with steadily decreasing orbital angular
momentum. Furthermore, initially non-resonant
systems can actually be captured into these stable orbits and then
oscillate around a fixed orientation throughout the rest of the
inspiral process. 

One effect of this resonance behavior is to significantly reduce
the size of the parameter space over which the spin orientations are
distributed. In turn, this could have an important impact on the size
of the template library used for waveform matching. Instead of
including a separate template for every single spin orientation (at
least four parameters), we could instead use a one- or two-dimensional
family of equilibrium solutions as representative of the entire
sample. These ``guiding center'' waveforms would be a good match for
any resonant system, librating around the equilibrium orientations,
potentially reducing the size of the search space by orders of
magnitude while maintaining a set of strictly physical templates.

Alternatively, we can take the inverse approach so that, given the
gravitational wave form of an inspiraling binary, we should be able to
infer the evolutionary history leading up to the point where it enters
the detector frequency band. In this manner, perturbative methods
could be used to explore the phase space around the equilibrium
regions and better pinpoint the exact binary parameters of the
detected signal. Eventually, as the detection rate increases to
hundreds or even thousands of signals per year, we will be able to
fully map out the spin distribution of compact binaries in the
Universe and thus test the theoretical results presented in this paper. 

In addition to reducing the volume of parameter space in the template
library, the equilibrium solutions could be extremely important for
determining initial conditions for the plunge calculations. This phase
of the binary merger is widely regarded as the least-well understood
physically and at the present time we have relatively little faith in any
theoretical waveforms that might describe the plunge (see, e.\ g.\
\citep{baker02} and references therein). Since the post-Newtonian expansion
(known at this point up to terms of order 3.5PN)
becomes less and less accurate in the plunging region, it is likely
that we will need to rely on full numerical relativity calculations to
produce the transitional waveforms linking the long inspiral and the
ring-down phase (which is itself well described by perturbative
methods). Because of the prohibitive computational expense of these
calculations, it is that much more important to have a solid understanding
of how to describe the initial conditions. We believe that the
equilibrium solutions described in this paper are an excellent place
to begin, as they may represent a large segment of the relevant binary
population. Knowing the final spin orientation at the plunge phase
would also improve our understanding of black hole recoil velocities,
a field of much recent interest \citep{favat04}.

In Section \ref{bin_evol} we present the evolution equations
and discuss the relevant time scales for the binary inspiral and spin
precession. In Section \ref{geometry_equil} we show how this system of
equations can be further reduced to five 
variables (four when assuming circular orbits), which completely
describe the relative orientation of the compact objects, and how this
reduced system exhibits resonance behavior. Section \ref{spin_evol}
describes the evolution of such a system under radiation reaction and
the dependence on initial conditions. In Sections \ref{mass_ratios}
and \ref{ecc_effects} we discuss the effects of large mass
ratios, eccentric orbits, and the dependence on spin magnitude. We
conclude with a discussion of the 
astrophysical implications of these results and directions for future
work.

\section{BINARY EVOLUTION EQUATIONS}\label{bin_evol}
We use the post-Newtonian equations of motion for the reduced two-body
problem including Lense-Thirring precession terms for the point mass
spins. Throughout this paper we adopt geometrized units with
$G=c=1$. The instantaneous precession
equations are (see, e.\ g.\ \citep{kidde93,apost94,kidde95})
\begin{subequations}
\begin{equation}\label{s1dot}
\mathbf{\dot{S}}_1 = \mathbf{\Omega}_1 \times \mathbf{S}_1
\end{equation}
\begin{equation}\label{s2dot}
\mathbf{\dot{S}}_2 = \mathbf{\Omega}_2 \times \mathbf{S}_2,
\end{equation}
\end{subequations}
where 
\begin{subequations}
\begin{equation}\label{omega_1}
\mathbf{\Omega}_1 = \frac{1}{r^3}\left
[ \left(2+\frac{3}{2}\frac{m_2}{m_1}\right)\mathbf{L}_N -\mathbf{S}_2
+3(\hat{\mathbf{r}}\cdot \mathbf{S}_2)\hat{\mathbf{r}}\right]
\end{equation}
\begin{equation}\label{omega_2}
\mathbf{\Omega}_2 = \frac{1}{r^3}\left
[ \left(2+\frac{3}{2}\frac{m_1}{m_2}\right)\mathbf{L}_N -\mathbf{S}_1
+3(\hat{\mathbf{r}}\cdot \mathbf{S}_1)\hat{\mathbf{r}}\right].
\end{equation}
\end{subequations}
Here $\mathbf{S}_{1,2}$ are the spin vectors of the two
compact objects with units of angular momentum and a dot represents time
derivative (assuming a global coordinate time $t$ in the
post-Newtonian approximation). The bodies have masses $m_1 \ge m_2$ and
are separated by a distance $r$ in the $\hat{\mathbf{r}}$
direction. The Newtonian orbital angular momentum $\mathbf{L}_N$ is
defined in the usual way: $\mathbf{L}_N = \mu (\mathbf{r}\times
\dot{\mathbf{r}})$ with the reduced mass defined as
\begin{equation}
\mu = \frac{m_1 m_2}{m}
\end{equation}
and the total mass is $m=m_1+m_2$.

At first glance, the leading $r^{-3}$ terms in equations
(\ref{omega_1},\ref{omega_2}) suggest that spin precession only
becomes important late in the inspiral when $r$ becomes small. However,
since the great majority of the evolution (in terms of time) occurs at a
large separation, the spins actually have a chance to undergo a large
number of slow precessions leading up to the inspiral phase. To
first order, the evolution of a circular orbit due to gravitational
radiation is \citep{peter64}
\begin{equation}
r(t) = (r^4(0) - \alpha t)^{1/4},
\end{equation}
where $\alpha = \frac{256}{5}\mu m^2$, giving  
\begin{equation}\label{dr_dt}
\frac{dr}{dt} = -\frac{\alpha}{4}(r^4(0)-\alpha t)^{-3/4}.
\end{equation}
We can thus define a characteristic time scale for inspiral as
\begin{equation}
T_{\rm inspiral} \equiv \frac{r}{dr/dt} \sim r^4.
\end{equation}
The precession time scale is given to leading order as
\begin{equation}
T_{\rm prec} \equiv |\Omega|^{-1} \sim \frac{r^3}{L_N} \sim
r^{5/2}.
\end{equation}
For large $r$, the precession time scale is actually shorter than the
inspiral time scale, and thus we see that the spin-orbit effects are
important even at early times. 

At these large separations, the system evolves very slowly in
time, with an orbital period much shorter than the precession
period. The number of orbits $N_{\rm orb}$ at a given separation $r$ and
orbital period $T_{\rm orb}$ is given by
\begin{equation}
\frac{dN_{\rm orb}}{dr} = \frac{1}{T_{\rm orb}}
\left(\frac{dr}{dt}\right)^{-1} \sim r^{3/2}
\end{equation}
while the number of precession cycles $N_{\rm prec}$ is given by
\begin{equation}
\frac{dN_{\rm prec}}{dr} = \frac{1}{T_{\rm prec}}
\left(\frac{dr}{dt}\right)^{-1} \sim r^{1/2}.
\end{equation}
In short, this means we must calculate the spin
evolution over a large range of $r$, corresponding to a very large
number of orbits. However, since we are interested here in the net
evolution of the spin orientation, the actual phase of the binary
orbit can be ignored and equations (\ref{omega_1}) and (\ref{omega_2}) can
be calculated in an orbit-averaged approximation (see
Appendix \ref{app_a}). Generalizing to non-circular orbits with
semimajor axis $a$ 
and eccentricity $e$, the orbit-averaged precession vectors are
\begin{subequations}
\begin{equation}\label{omg1avg}
\mathbf{\bar{\Omega}}_1 = \frac{1}{a^3(1-e^2)^{3/2}}
\left[\left(2+\frac{3}{2}\frac{m_2}{m_1}-\frac{3}{2}\frac{\mathbf{S}_2\cdot
\mathbf{L}_N}{L_N^2}\right)\mathbf{L}_N +\frac{1}{2}\mathbf{S}_2\right]
\end{equation}
and
\begin{equation}\label{omg2avg}
\mathbf{\bar{\Omega}}_2 = \frac{1}{a^3(1-e^2)^{3/2}}
\left[\left(2+\frac{3}{2}\frac{m_1}{m_2}-\frac{3}{2}\frac{\mathbf{S}_1\cdot
\mathbf{L}_N}{L_N^2}\right)\mathbf{L}_N +\frac{1}{2}\mathbf{S}_1\right].
\end{equation}
\end{subequations}

An interesting result is that the orbit-averaged precession equations
are independent of the ellipse's orientation in the plane normal to
$\mathbf{L}_N$. In other words, the first-order post-Newtonian (1PN)
effect of pericenter precession does not play a role in the spin
evolution, which is fortunate because of the relatively short time
scales involved, compared to the slower effects of spin-orbit
acceleration (1.5PN), spin-spin acceleration (2PN), and radiation
reaction (2.5PN) \citep{kidde93,kidde95}. In the orbit-averaged equations of
motion, the spin-orbit
precession terms (1PN) dominate on a short time scale, but the secular
effects of the radiation reaction are also very important in the
long-term evolution of the system.

On time scales short compared to the inspiral time, the total angular
momentum 
\begin{equation}\label{Jtot}
\mathbf{J} =\mathbf{L}_N+\mathbf{S}=
\mathbf{L}_N+\mathbf{S}_1+\mathbf{S}_2 
\end{equation}
is conserved, implying that the orbital angular momentum evolves
according to 
\begin{equation}\label{spinldot}
\dot{\mathbf{L}}_N = -\dot{\mathbf{S}} \sim \frac{1}{r^3}\mathbf{S}_{\rm eff}
\times \mathbf{L}_N,
\end{equation}
where $\mathbf{S}_{\rm eff}$ is some linear combination of $\mathbf{S}_1$
and $\mathbf{S}_2$ (note this is \textit{not} the same effective spin
as in \citep{buona04}).
Since $\mathbf{L}_N\cdot \dot{\mathbf{L}}_N=0$, without radiation
reaction, the magnitude of the orbital angular 
momentum vector is constant with spin precession. The magnitude of the
total spin vector $\mathbf{S}$, on the other hand, is \textit{not}
conserved as the angle changes between the two spin vectors (each of
constant magnitude). 

The relations (\ref{Jtot}) and (\ref{spinldot}) constrain the
binary system to a subset of the complete parameter space defined by
the three vectors $\mathbf{L}_N$, $\mathbf{S}_1$, and
$\mathbf{S}_2$. We believe it is this set of constraints that best
explains much of the behavior presented below, as opposed to a more
classical description of resonance based on Hamiltonian mechanics and
energy minima in phase space (see, e.\ g.\
\citet{murra99,sussm01}). However, a Hamiltonian formulation of the
post-Newtonian equations of motion such as in \citep{damou88,jaran98,damou01}
may prove to give a more classical explanation to these apparently
geometric constraints. 

The inclusion of gravitational radiation causes the orbit to shrink and also
circularize in time, reducing $a$, $e$, and the magnitude of the
angular momentum
\begin{equation}\label{L_ae}
L_N = \mu \sqrt{ma(1-e^2)}.
\end{equation}
Following \citet{peter64} and adopting units with $m=1$, we use the
coupled first order differential equations
\begin{equation}\label{da_dt}
\frac{d}{dt}a = -\frac{64}{5}\frac{\mu}{a^3(1-e^2)^{7/2}}
\left(1+\frac{73}{24}e^2+\frac{37}{96}e^4\right), 
\end{equation}
\begin{equation}\label{de_dt}
\frac{d}{dt}e = -\frac{304}{15}\frac{\mu e}{a^4(1-e^2)^{5/2}}
\left(1+\frac{121}{304}e^2\right), 
\end{equation}
and
\begin{equation}\label{dL_dt}
\frac{d}{dt}\mathbf{L}_N = -\frac{32}{5}\frac{\mu^2}{a^{7/2}(1-e^2)^2}
\left(1+\frac{7}{8}e^2\right) \hat{\mathbf{L}}_N
\end{equation}
to evolve the binary orbital elements in time. All of the above
orbit-averaged precession and radiation reaction
equations have been tested and compared to the full 2.5-order
post-Newtonian equations of motion in \citet{kidde95}. The agreement is
very good for most of the inspiral, all the way down to $r \lesssim
10m$, after which almost any post-Newtonian approximation becomes
increasingly uncertain.

\section{GEOMETRY OF EQUILIBRIUM}\label{geometry_equil}
\begin{figure}[bp]
\caption{\label{schematic} Schematic diagram of the spin and orbital
  angular momentum vectors. The coordinate system is defined such that
  $\mathbf{L}_N$ is along the z-axis and $(\theta_1,\theta_2)$ are
  the respective angles between $\mathbf{L}_N$ and $(\mathbf{S}_1,
  \mathbf{S}_2)$. The projection of $\mathbf{S}_1$ onto the x-y plane
  is defined to be along the x-axis so the azimuthal spin angles are
  $\phi_1 = 0$ and $\phi_2 = \Delta\phi$.}
\begin{center}
\scalebox{1.5}{\includegraphics{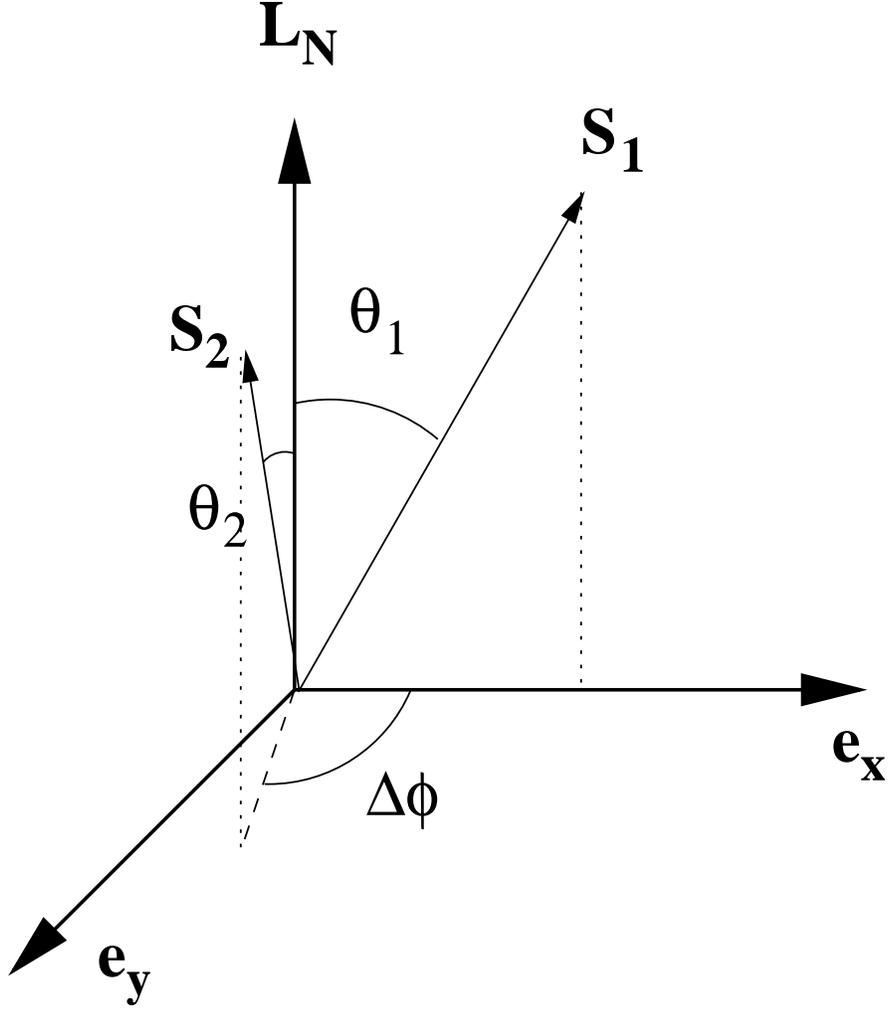}}
\end{center}
\end{figure}

One of the most difficult aspects of studying the spinning binary
system is the problem of visualizing and analyzing the orientation of
the two spins and the angular momentum in an informative way. In
general, these three vectors are defined by nine coordinates [the
angular momentum is also related to $a$ and $e$ through
(\ref{L_ae})]. Since the spin magnitudes $S_1$ and $S_2$ are conserved
in the point mass approximation, and we can pick a coordinate system
where $\mathbf{L}_N$ points in the $\hat{\mathbf{e}}_z$ direction, we
are left with five coordinates: $(L_N, \theta_1, \theta_2, \phi_1,
\phi_2)$. Furthermore, the overall dynamics are preserved under
rotation around $\mathbf{L}_N$ so we can reduce the spin degrees of
freedom by defining the $\hat{\mathbf{e}}_x$ direction along
$\phi_1=0$, leaving four independent coordinates to define the
orientation of the system: $(L_N, \theta_1, \theta_2,
\Delta\phi)$. Figure \ref{schematic} shows a schematic of the geometry
used throughout this paper. Following the post-Newtonian formalism,
all angles and vector magnitudes are defined in a Cartesian, flat
space-time. 

In this coordinate system, there exists a set of equilibrium spin
configurations for which $L_N, \theta_1, \theta_2,$ and
$\Delta\phi$ are constant (without radiation reaction), even though
the individual vectors might vary in time from the
point of view of a fixed inertial coordinate system. Trivial
equilibrium examples include the collinear cases with
$\cos\theta_1 = \pm 1$ and $\cos\theta_2 = \pm 1$. More interesting
cases occur when $\mathbf{S}_1$, $\mathbf{S}_2$, and $\mathbf{L}_N$
all appear to precess around a fixed axis at a constant rate so as to
remain in a fixed relative orientation. These points in parameter
space can be found by solving (cf.\ \citet{apost96})
\begin{equation}\label{s1dots2}
\frac{d}{dt}(\mathbf{S}_1 \cdot \mathbf{S}_2) =
\frac{3}{2a^3(1-e^2)^{3/2}}\left[\frac{m_2}{m_1}-\frac{m_1}{m_2}+
  \frac{(\mathbf{S}_1-\mathbf{S}_2)\cdot\mathbf{L}_N}{L^2_N}\right]
\mathbf{S}_2\cdot(\mathbf{L}_N\times\mathbf{S}_1) = 0.
\end{equation}
Note that the last term (a vector triple-product) can be written in
our reduced coordinate system as
\begin{equation}
\mathbf{S}_2\cdot(\mathbf{L}_N\times\mathbf{S}_1) =
S_1S_2L_N\sin\theta_1\sin\theta_2\sin\Delta\phi.
\end{equation}
Thus (\ref{s1dots2}) is satisfied when $\mathbf{S}_1$, $\mathbf{S}_2$,
and $\mathbf{L}_N$ are coplanar, i.e. $\sin\Delta\phi = 0$ for all
times. In practice this means finding simultaneous solutions to
\begin{equation}
\mathbf{S}_2\cdot(\mathbf{L}_N \times \mathbf{S}_1) = 0
\end{equation}
and
\begin{equation}\label{ddts2Ls1}
\frac{d}{dt}[\mathbf{S}_2\cdot(\mathbf{L}_N \times \mathbf{S}_1)]=0. 
\end{equation}
Combining with equations (\ref{omega_1}, \ref{omega_2}, \ref{omg1avg}, and
\ref{omg2avg}), (\ref{ddts2Ls1}) can be written in terms of just
$\mathbf{S}_1, \mathbf{S}_2,$ and $\mathbf{L}_N$, with no explicit
time derivatives:
\begin{equation}\label{equilib}
(\mathbf{\Omega}_1\times\mathbf{S}_1)\cdot
  [\mathbf{S}_2\times(\mathbf{L}_N + \mathbf{S}_1)] =
(\mathbf{\Omega}_2\times\mathbf{S}_2)\cdot
  [\mathbf{S}_1\times(\mathbf{L}_N + \mathbf{S}_2)],
\end{equation}
which in turn are simple combinations of our reduced coordinates
$(L_N, \theta_1, \theta_2, \Delta\phi)$.

\begin{figure}[bp]
\caption{\label{equ_align} Equilibrium solutions of equation
  (\ref{equilib}) for the case of maximally spinning black holes with
  normalized masses $(m_1=0.55, m_2 = 0.45)$ and spin alignment
  $\Delta\phi = 0^\circ$. The solution curves are labeled by their different
  values of the orbital angular momentum $L_N$, so that equilibrium
systems evolving under radiation reaction move along subsequent curves
with decreasing $L_N$ (corresponding to $r/m \approx 260,150,65,16,4$). The
labels 'A' and 'B' refer to the initial conditions for Figures
\ref{dphitot}a and \ref{dphitot}b, both with $L_N(t_0)=4$.}
\begin{center}
\scalebox{0.9}{\includegraphics{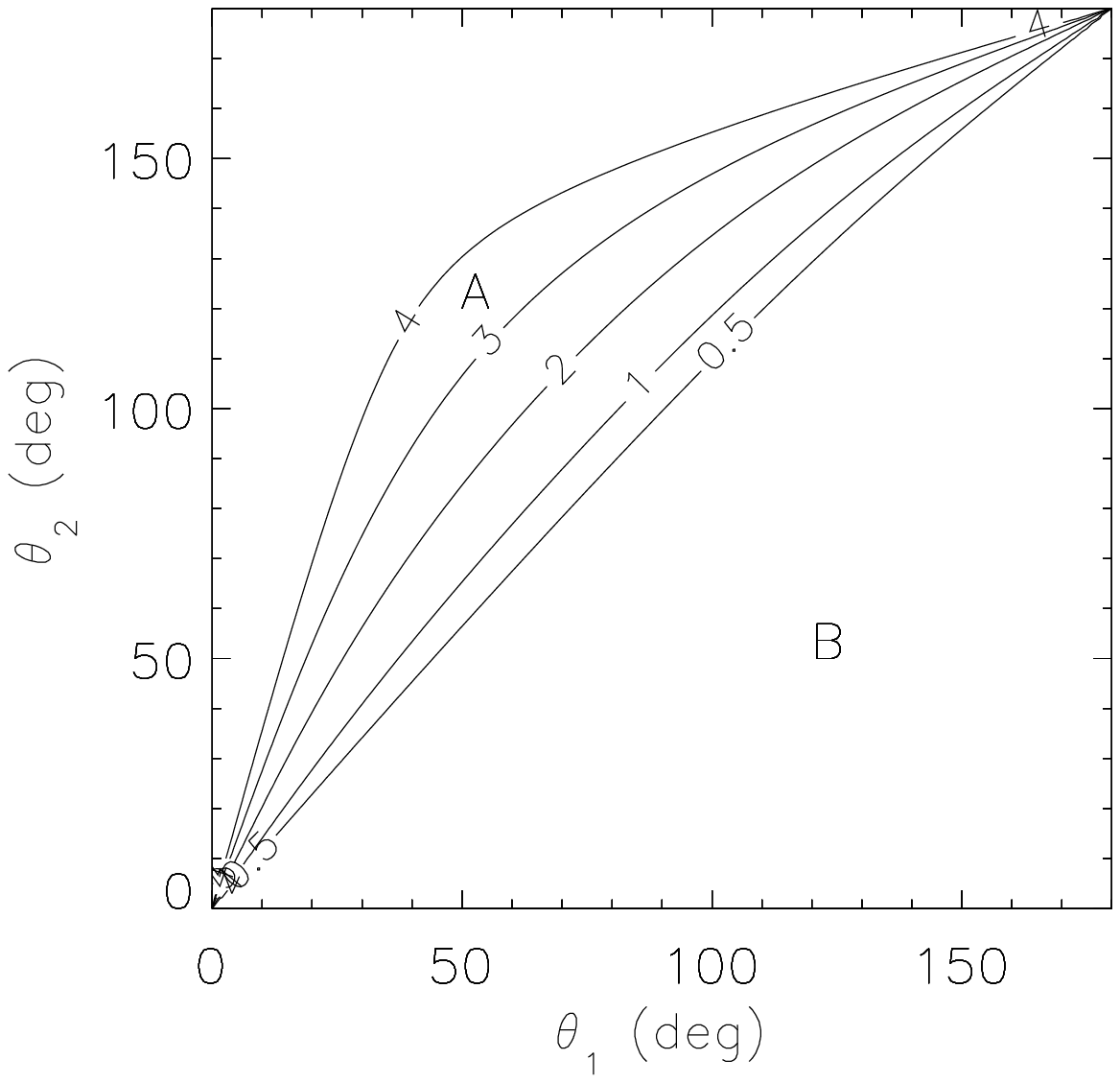}}
\end{center}
\end{figure}

\begin{figure}[bp]
\caption{\label{equ_anti} Equilibrium solutions of equation
  (\ref{equilib}) for the case of maximally spinning black holes with
  normalized masses $(m_1=0.55, m_2 = 0.45)$ and spin anti-alignment
  $\Delta\phi=180^\circ$. The solution curves are labeled by their different
  values of the orbital angular momentum $L_N$, so that equilibrium
systems evolving under radiation reaction move along subsequent curves
with decreasing $L_N$ (corresponding to $r/m \approx 260,150,65,16,4$). Solid
curves correspond to stable equilibrium orientations, while the dashed
curves are the quasi-stable solutions described in the text. The
label 'C' refers to the unstable initial conditions of Figure
\ref{dphitot}c, with $L_N(t_0)=1$.}
\begin{center}
\scalebox{0.9}{\includegraphics{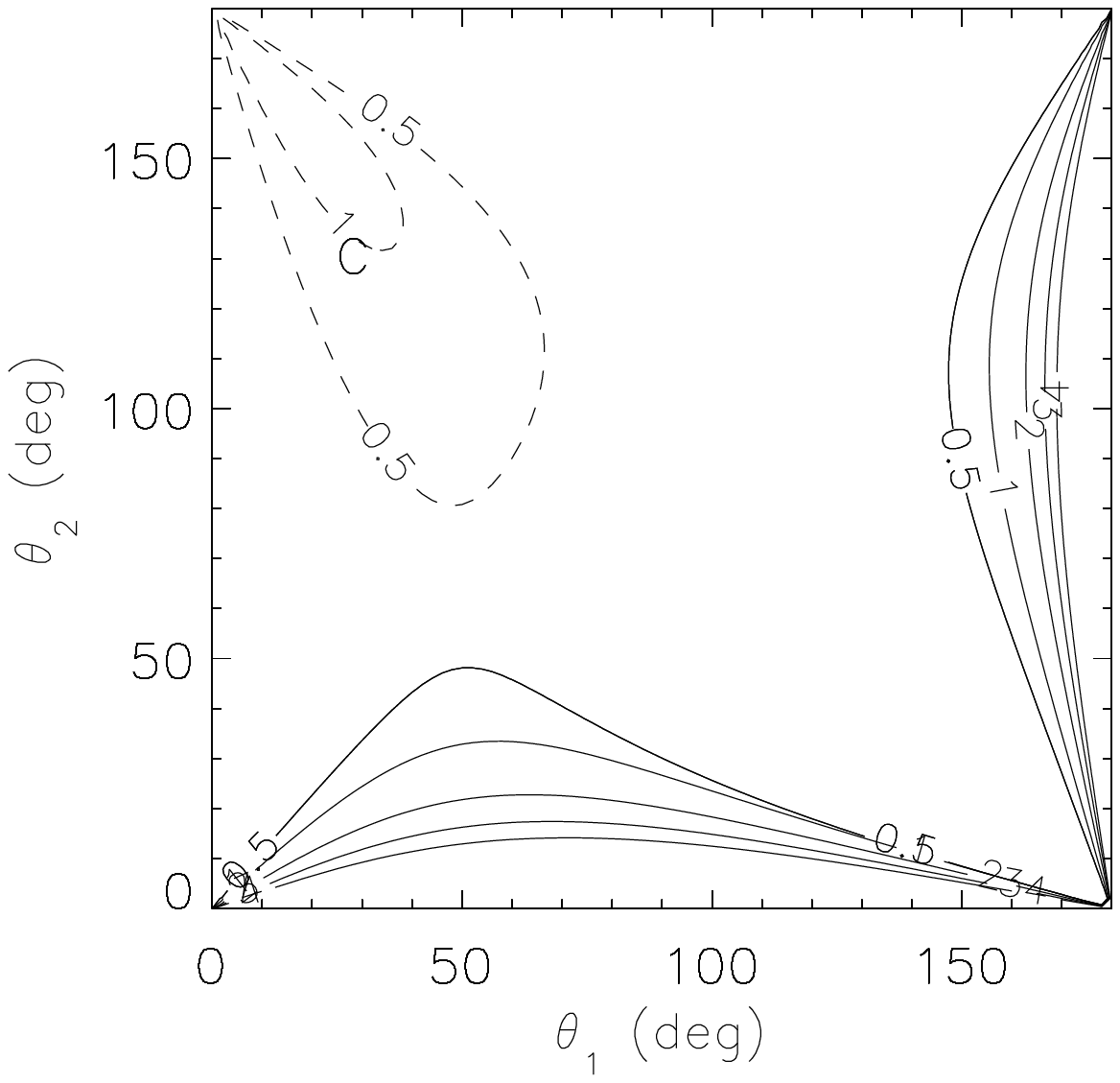}}
\end{center}
\end{figure}

For a given value of $L_N$ (i.\ e.\ at a particular point in time
during the binary inspiral), and setting $\sin \Delta \phi = 0$,
solutions of (\ref{equilib}) trace out
one-dimensional curves in $(\theta_1,\theta_2)$ space. Figure
\ref{equ_align} shows these curves for two maximally spinning black
holes with nearly equal masses $(m_1 = 0.55, m_2=0.45)$ at a few
stages of the inspiral evolution ($L_N = 4, 3, 2, 1, 0.5$,
with binary separations of $r/m \approx 260, 150, 65, 16, 4$). These
solutions correspond to the aligned configuration with
$\Delta\phi=0$ and are found numerically to be stable, as will be
explained below. There also exist anti-aligned equilibrium curves for
$\Delta\phi = 180^\circ$, shown in 
Figure \ref{equ_anti} for the same masses and values of $L_N$. We
find that the curves along the bottom $(\theta_2 \lesssim 50^\circ)$ and the
right side $(\theta_1 \gtrsim 140^\circ)$ of the plot correspond to stable
equilibriums, while the solutions in the upper left corner, appearing at
later times, are generally unstable or quasi-stable.

The regions of stability can be better understood by considering a
type of effective potential for orbits in parameter space near the
equilibrium solutions. While the precession equations
do not suggest any obvious way of being
reformulated in the traditional approach of an analytic effective
potential, we can map out the potential numerically by
integrating a large sample of initial conditions. For each of these
initial conditions, the stability of the spin orientation can be
measured by integrating the change in $\Delta\phi$:
\begin{equation}
[\Delta\phi(t)]_{\rm tot} = \int_{t_0}^t \frac{d}{dt'}\Delta\phi(t') dt'.
\end{equation}
With this definition, $[\Delta\phi(t)]_{\rm tot}$ is potentially unbounded
and can grow in magnitude larger than $360^\circ$ as one spin precesses around
$\mathbf{L}_N$ more rapidly and repeatedly ``passes'' the other spin
vector.
For binaries locked in a spin equilibrium (with either $\Delta\phi=0^\circ$
or $\Delta\phi=180^\circ$), the integrated phase shift will be zero. For
systems far from this equilibrium, the two spins will precess more or
less independently and will generally acquire a linear phase shift in
time since $|\bar{\Omega}_1| \neq |\bar{\Omega}_2|$. And for
systems \textit{near} the equilibrium points in phase space, we expect
the phase to librate around the exact coplanar solution given by
equation (\ref{equilib}). 

\begin{figure}[bp]
\caption{\label{dphitot} Integrated phase difference
  $[\Delta\phi(t)]_{\rm tot}$ between two precessing spin
  vectors. Examples are shown for initial conditions (a)
  near a stable equilibrium with $\theta_1=50^\circ, \theta_2=120^\circ,
  \Delta\phi=0^\circ,$ and $L_N = 4$; (b) far away from equilibrium with
  $\theta_1=120^\circ, \theta_2=50^\circ, \Delta\phi=0^\circ,$ and
$L_N = 4$; (c) near a quasi-stable equilibrium with $\theta_1=30^\circ,
\theta_2=133^\circ, \Delta\phi=180^\circ,$ and $L_N = 1$ (compare with
Figures \ref{equ_align} and \ref{equ_anti}). For all three cases,
radiation reaction has been turned off.} 
\begin{center}
\scalebox{0.9}{\includegraphics{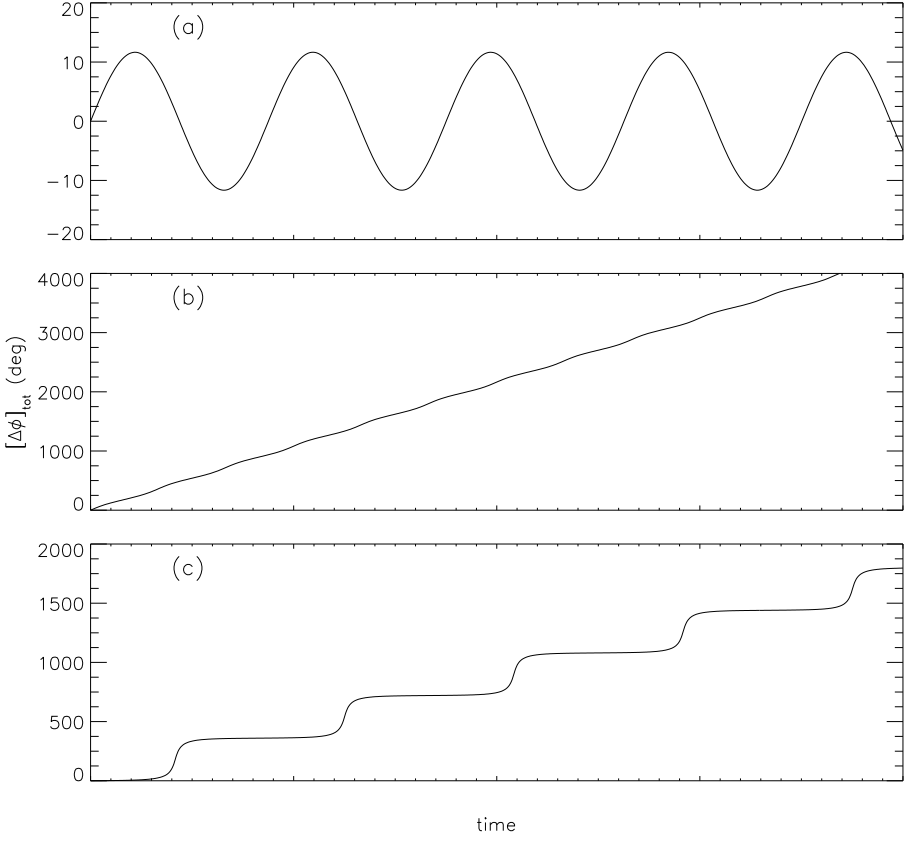}}
\end{center}
\end{figure}

Figure \ref{dphitot} shows 
$[\Delta\phi(t)]_{\rm tot}$ for a few examples of these different regions
in parameter space, again with $m_1 = 0.55, m_2 =0.45$, and maximum
spin parameters ($S_j=m_j^2$). Using Figures \ref{equ_align} and
\ref{equ_anti}
as guides, we can identify initial conditions near stable and unstable
equilibrium solutions. Figure \ref{dphitot}a shows a binary near a stable
equilibrium with $\theta_1=50^\circ, \theta_2=120^\circ, \Delta\phi=0,$ and
$L_N = 4$. The system librates around $\Delta\phi=0$ with amplitude of
about $11^\circ$. For initial conditions far away from equilibrium points,
the two spins precess at independent rates, giving a roughly linear
drift in phase, as seen in Figure \ref{dphitot}b, which has initial
conditions $\theta_1=120^\circ, \theta_2=50^\circ, \Delta\phi=0,$ and
$L_N = 4$. These two examples are also labeled on Figure
\ref{equ_align} as 'A' and 'B'.

For the systems oscillating around a stable equilibrium, we can think of
an effective potential whose form can be inferred by the
libration of trajectories through those points in phase space. With
this approach, the minima of the potential
lie along the solid curves plotted in Figures \ref{equ_align} and
\ref{equ_anti}. Systems lying near these curves in parameter space
will oscillate around them with constant amplitude and
frequency. Phase space trajectories like the one in Figure
\ref{dphitot}b correspond to a plateau-type region of the effective
potential with constant amplitude, where the spin vectors can
precess freely through all values of $\Delta\phi$.

From this analysis, we can also determine that some of the equilibrium
solutions plotted in Figure \ref{equ_anti} are unstable (dashed lines).
Perhaps it is more accurate to call these
regions \textit{quasi-stable} since initial conditions a small
distance outside of these closed curves are found to exhibit behavior
typical of trajectories near an unstable equilibrium: long periods of
stasis followed by short bursts of rapid divergence away from
equilibrium, as in Figure \ref{dphitot}c ($\theta_1=30^\circ,
\theta_2=133^\circ,\Delta\phi=180^\circ,L_N=1$, labeled 'C' on Figure
\ref{equ_anti}). However, points just
\textit{inside} the equilibrium curves appear stable, librating around
the locked $\Delta\phi=180^\circ$ position in phase space. 

In practice, we find that binaries initially outside of these
quasi-stable regions
tend to remain outside. Furthermore, when including radiation
reaction, \textit{all} binaries start outside of these regions, which
only begin to appear for relatively small values of $L_N$ late in the
evolution. Thus it would be quite unlikely
for an astrophysically realistic binary system to be found
on the stable side of any of the quasi-stable equilibrium
curves. Yet these regions are still of significant physical interest,
as we believe the boundaries between stable and
unstable regions in parameter space may be relevant to the ongoing
question of chaos in spinning compact binaries and the existence of
positive Lyapunov exponents in such systems (see, e.\ g.\
\citep{schni01,corni03}). 

\section{SPIN EVOLUTION}\label{spin_evol}
In Section \ref{geometry_equil}, we assumed in the analysis that the
orbital angular 
momentum $L_N$ was constant in magnitude, i.\ e.\ there was no radiation
reaction damping the system through gravitational wave emission. Of
course, in all physically realistic black hole binaries, gravitational
radiation will play a major role in the secular evolution of the
orbit. From the post-Newtonian analysis in Section \ref{bin_evol}, we
see that both radiation
reaction and spin precession are dynamically important through most of
the long inspiral process.

With the inclusion of radiation reaction, the angular momentum evolves
according to equations (\ref{spinldot}) and (\ref{dL_dt}), changing
direction due to Lense-Thirring precession and losing magnitude due to
gravitational losses. The individual spin vectors still maintain
constant magnitude under radiation reaction, but have access to a
broader range of precession angles since the total angular momentum
$\mathbf{J}$ is no longer conserved. In this sense, radiation reaction
removes an integral of
motion from the precession equations, allowing a greater region of
phase space to be sampled throughout the inspiral evolution. 

For the stable equilibrium solutions to equation (\ref{equilib}), we
find that the binary
systems that lie on or sufficiently close to one of the curves in Figures
\ref{equ_align} and \ref{equ_anti} will evolve along successive curves
of decreasing $L_N$ as the system loses angular momentum to radiation
reaction. At the same time, these stability regions grow stronger,
confining a wider range of spins to the equilibrium curves. Even for some
initial conditions that begin far away from the stable solutions,
precessing freely through a large range of $\Delta\phi$, the
effect of radiation reaction causes the spins to become locked and
then librate around equilibrium. These systems undergo
transitions from the ``drifting'' behavior of Figure \ref{dphitot}b to
the locked condition of \ref{dphitot}a. Once the system
becomes locked, we find it will stay locked throughout the rest of the
inspiral.

\begin{figure}[bp]
\caption{\label{sinalgn} Sinusoidal projections of spin parameter space
  $[\theta_{12}\equiv
\cos^{-1}(\hat{\mathbf{S}}_1\cdot\hat{\mathbf{S}}_2), \Delta\phi]$. A
sinusoidal
projection maps spherical coordinates $(\theta,\phi)$ to flat
Cartesian coordinates $(\phi\sin\theta,\theta)$ with equal solid
angles mapping to equal areas.
  The snapshots show a random sample of initial
  conditions with $r(t_0)=1000m$ evolving under radiation reaction at
  times when $r/m = 1000,250,125,10$. The masses of the two compact
objects are similar, with $m_1=0.55$ and $m_2=0.45$, and both have
maximal spins. The
initial spin of the larger mass is closely aligned with the orbital
angular momentum: $\theta_1(t_0) = 10^\circ$.} 
\begin{center}
\scalebox{0.4}{\includegraphics*[0,400][584,680]{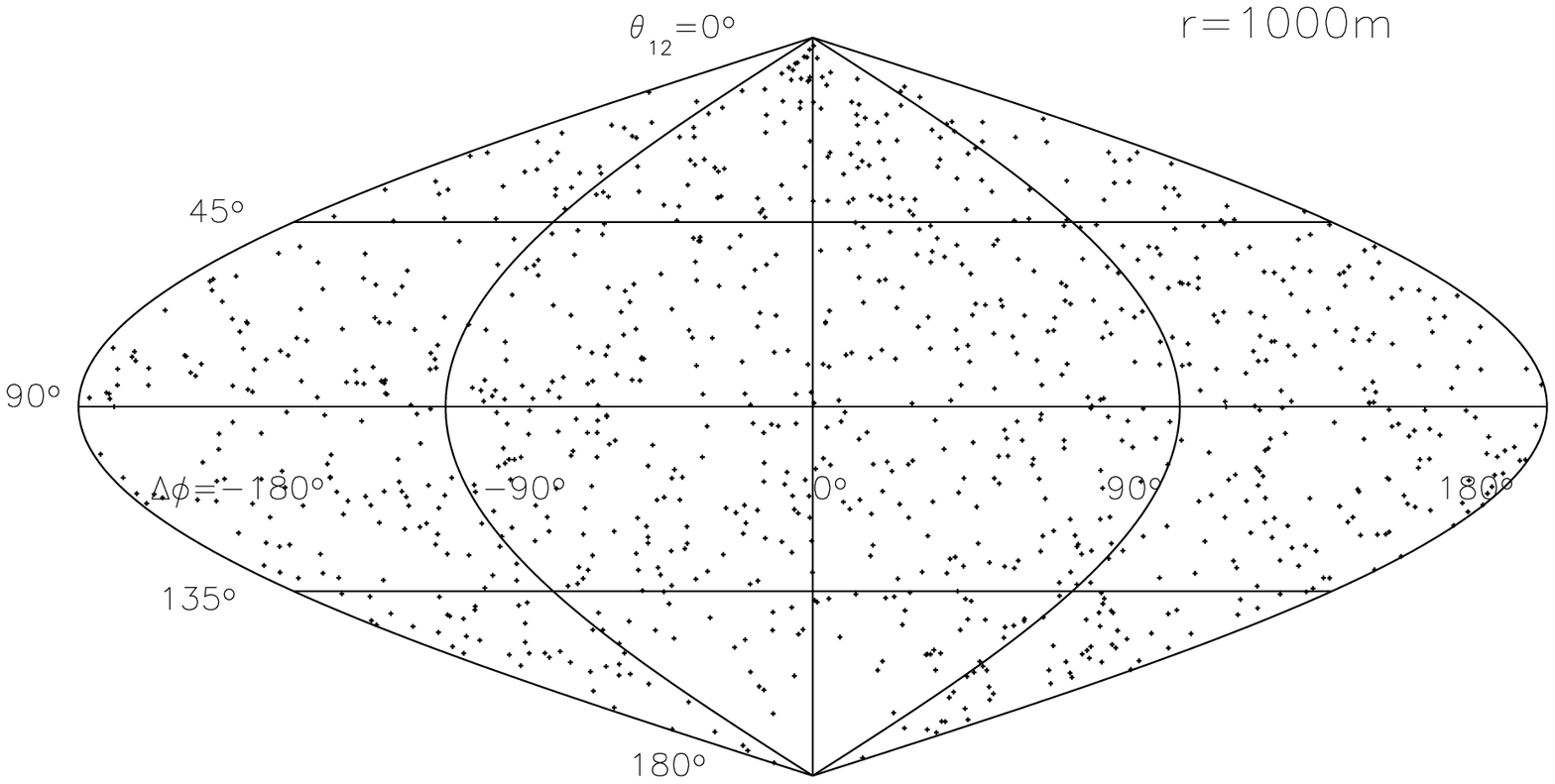}}
\scalebox{0.4}{\includegraphics*[0,400][584,680]{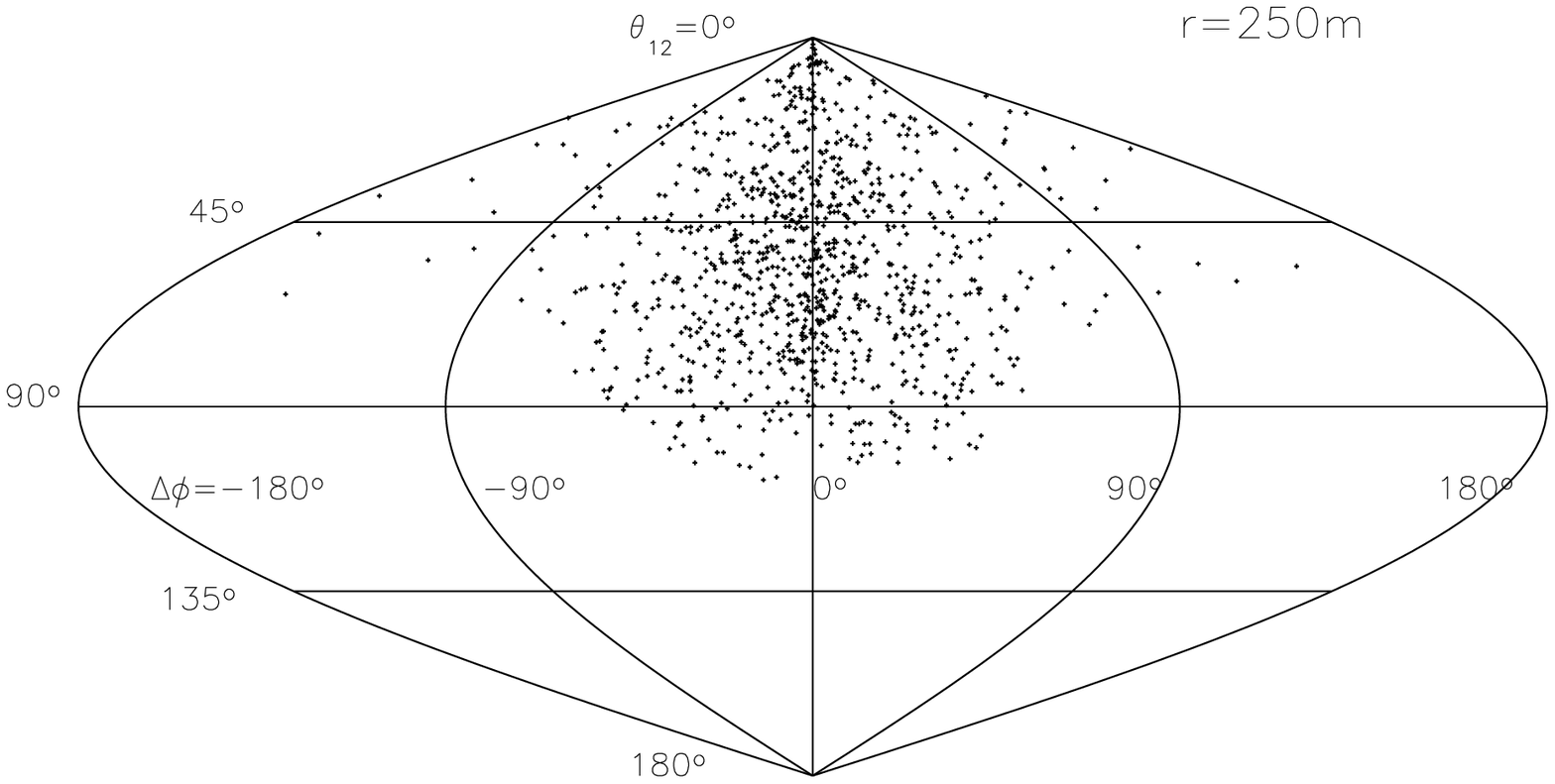}}
\scalebox{0.4}{\includegraphics*[0,400][584,680]{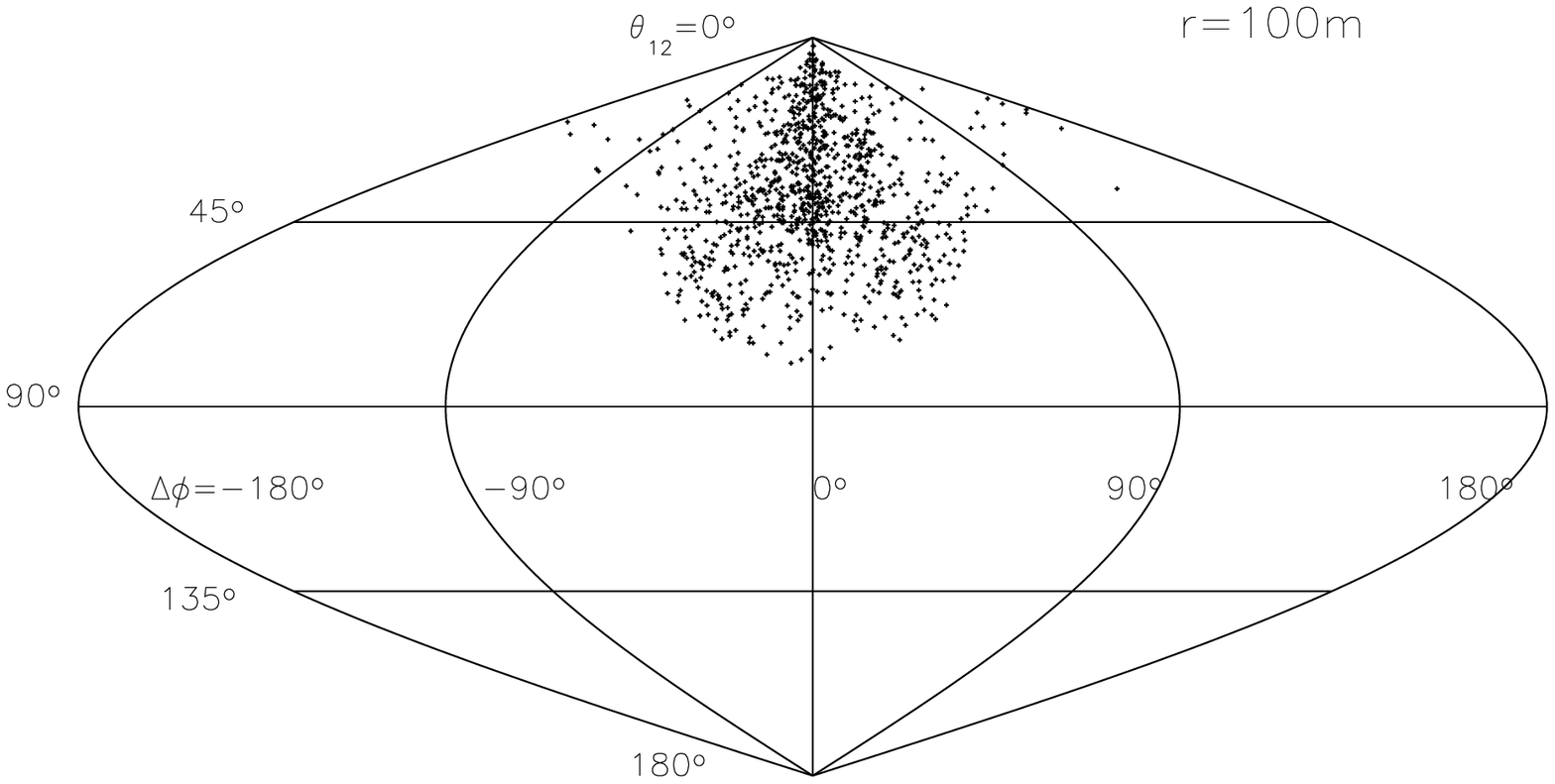}}
\scalebox{0.4}{\includegraphics*[0,400][584,680]{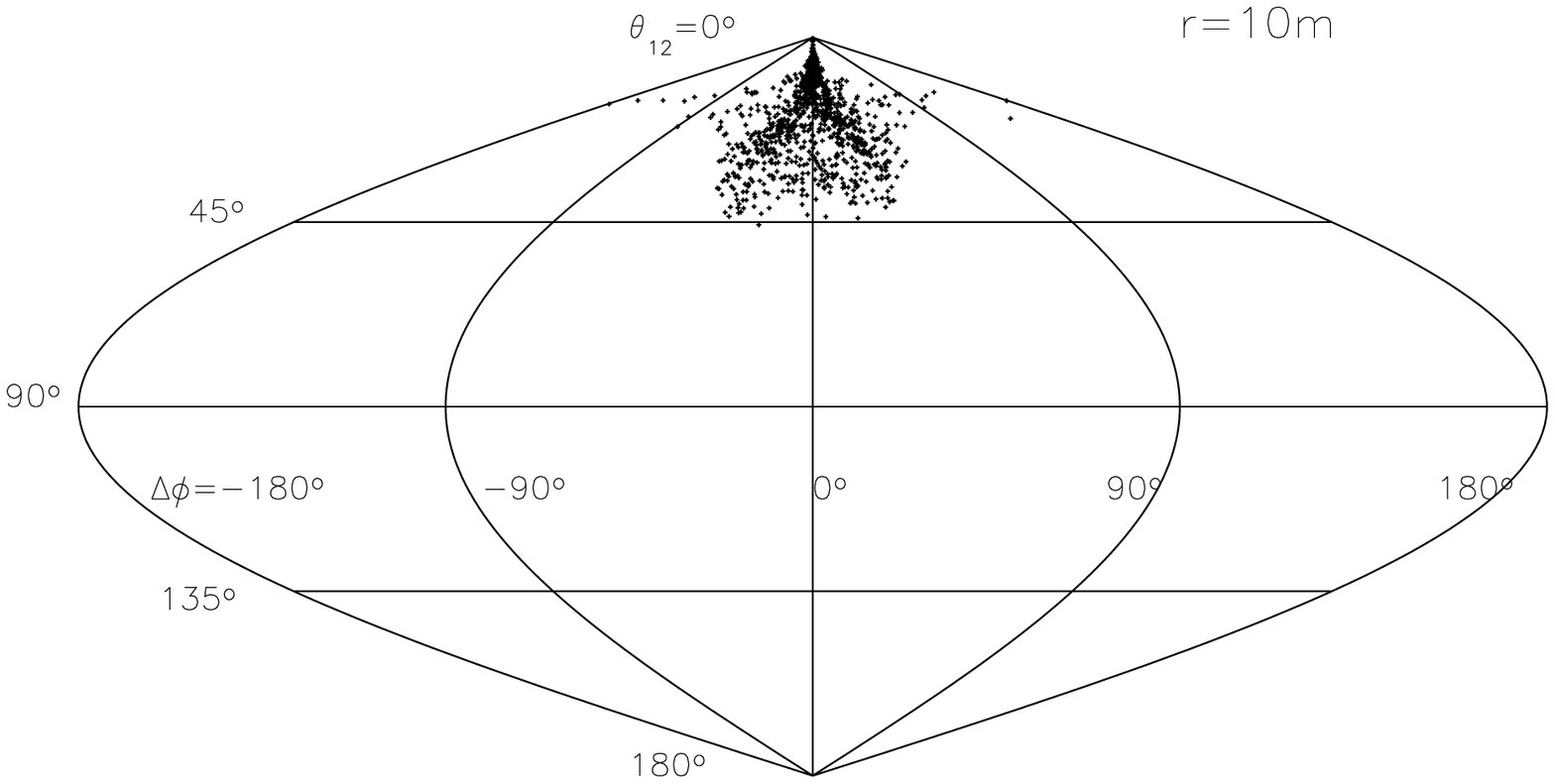}}
\end{center}
\end{figure}

\begin{figure}[bp]
\caption{\label{sinanti} Sinusoidal projections of spin parameter space
  $[\theta_{12}\equiv \cos^{-1}(\hat{\mathbf{S}}_1\cdot\hat{\mathbf{S}}_2),
  \Delta\phi]$, as in Figure \ref{sinalgn}. The snapshots show a
random sample of initial
  conditions with $r(t_0)=1000m$ evolving under radiation reaction at
  times when $r/m = 1000,250,125,10$. The masses of the two compact
objects are similar, with $m_1=0.55$ and $m_2=0.45$, and both have
maximal spins. The
initial spin of the larger mass is misaligned with the orbital angular
momentum: $\theta_1(t_0) = 170^\circ$.}
\begin{center}
\scalebox{0.4}{\includegraphics*[0,400][584,680]{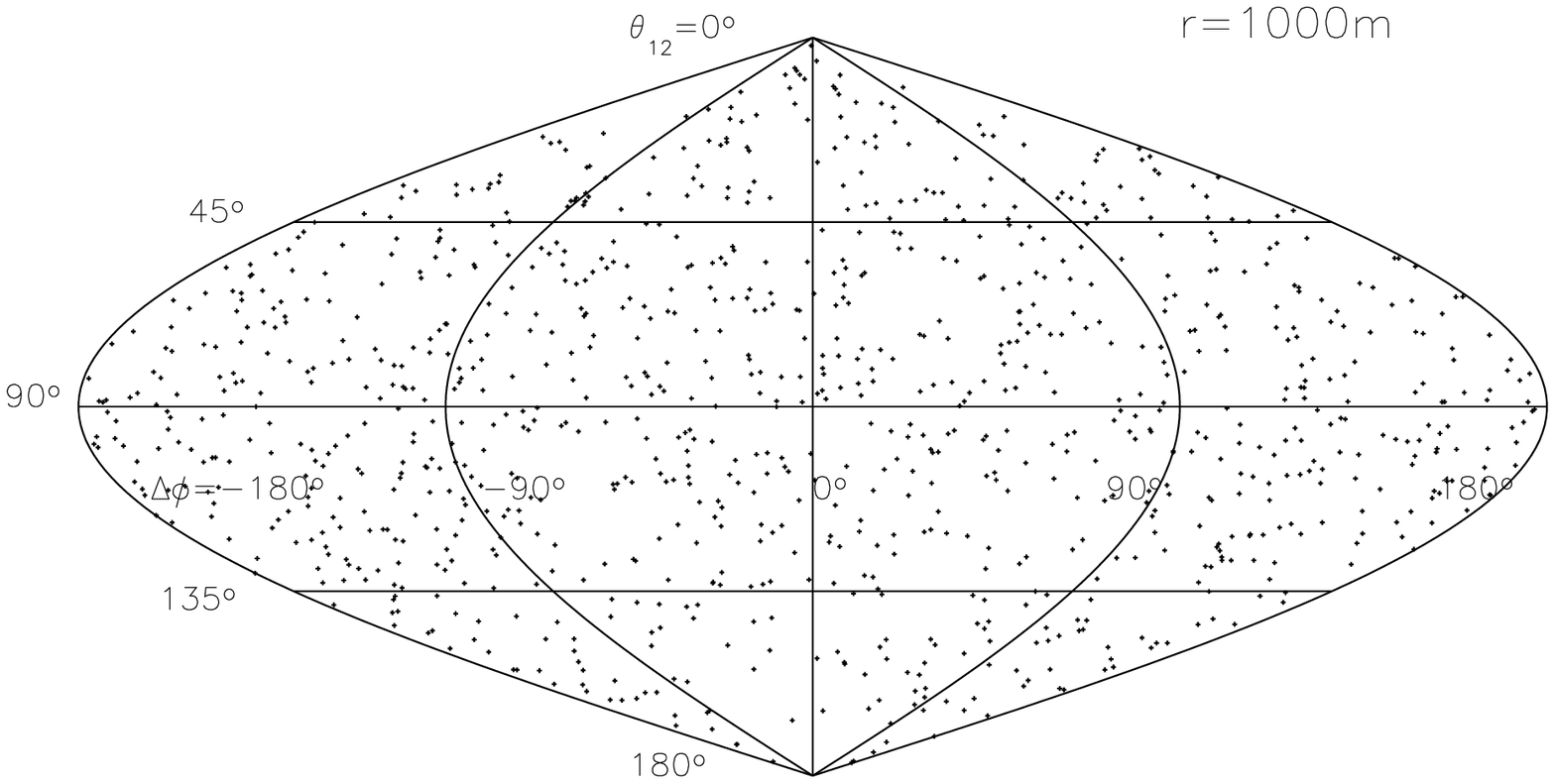}}
\scalebox{0.4}{\includegraphics*[0,400][584,680]{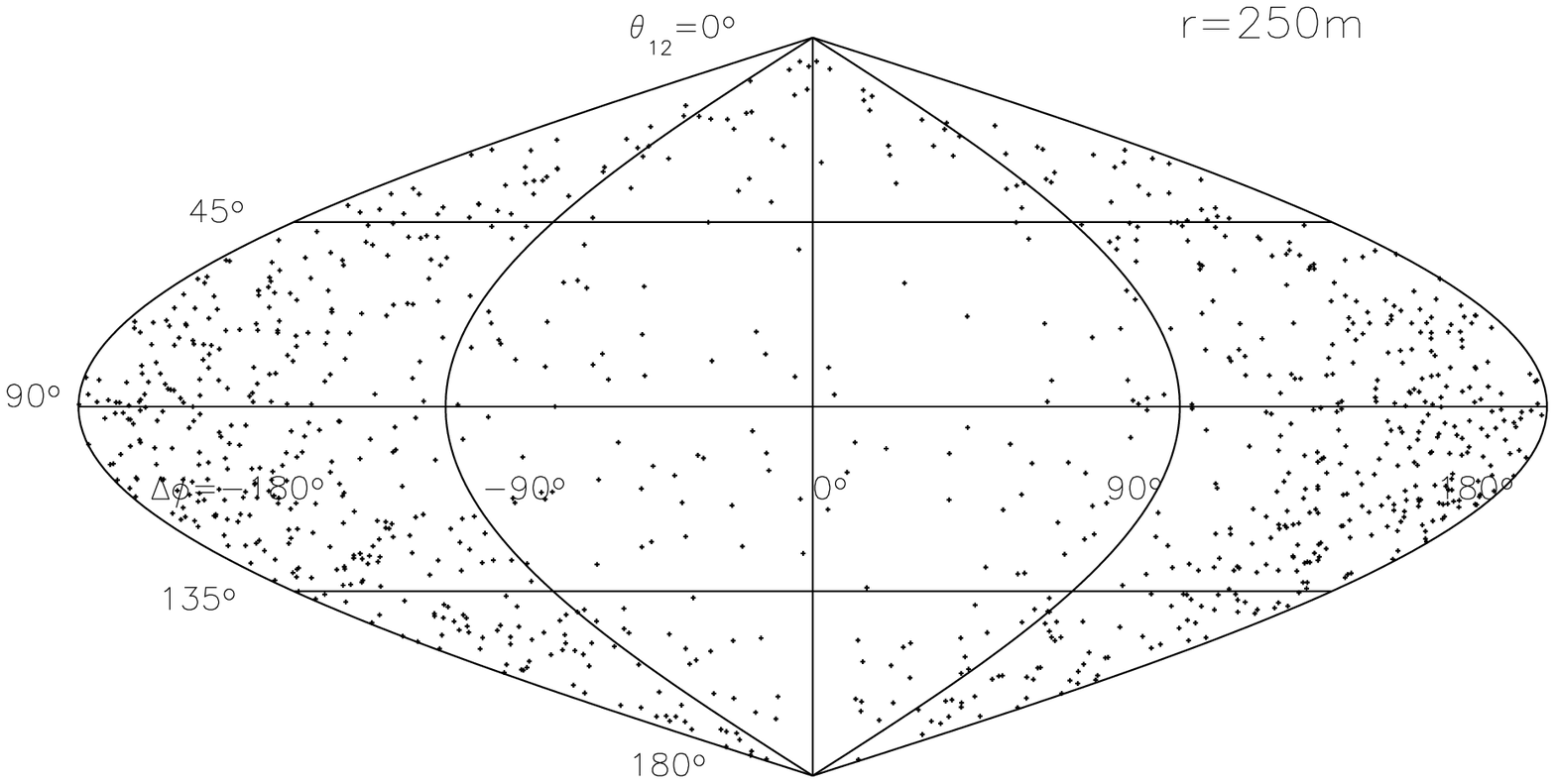}}
\scalebox{0.4}{\includegraphics*[0,400][584,680]{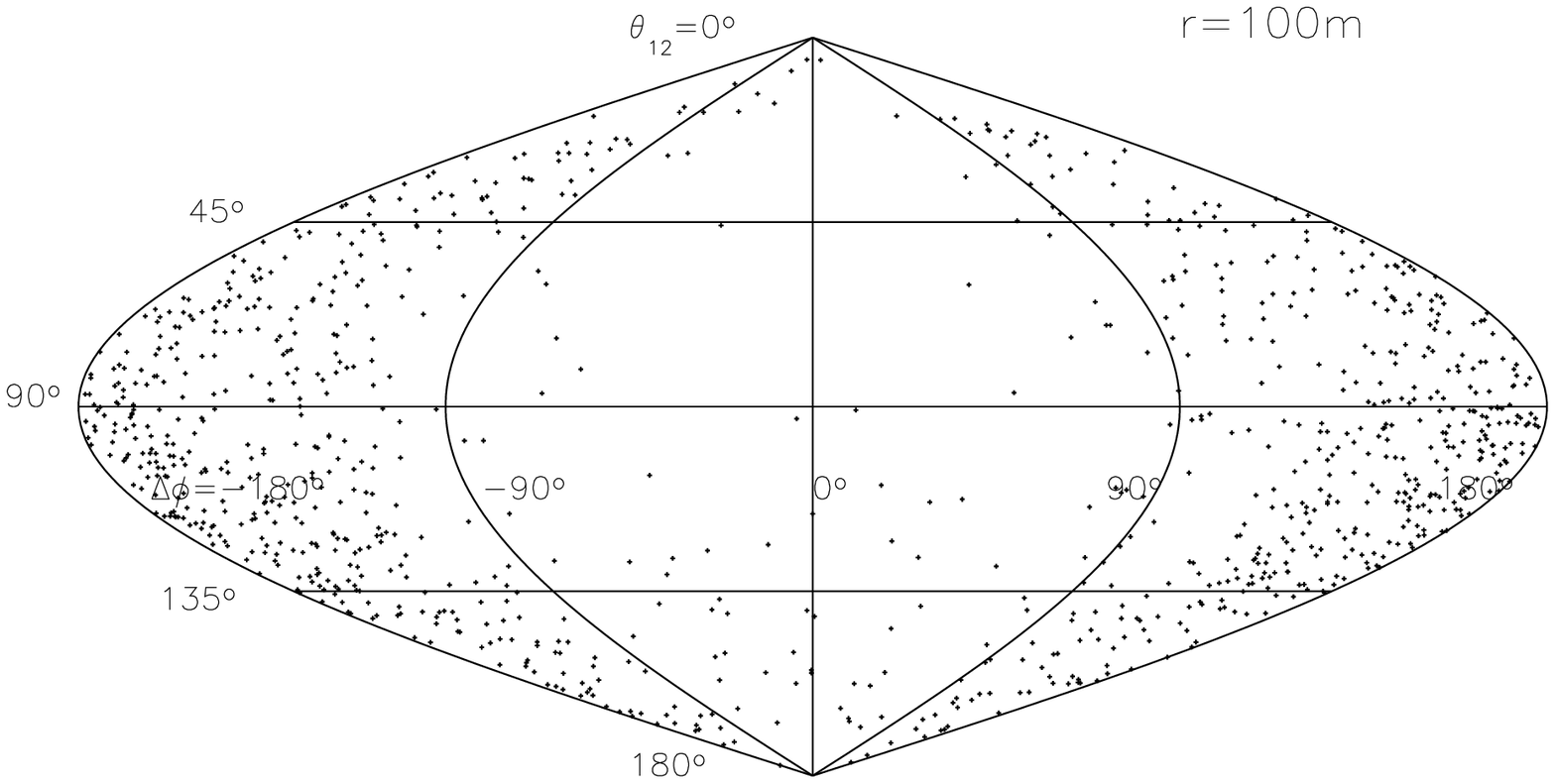}}
\scalebox{0.4}{\includegraphics*[0,400][584,680]{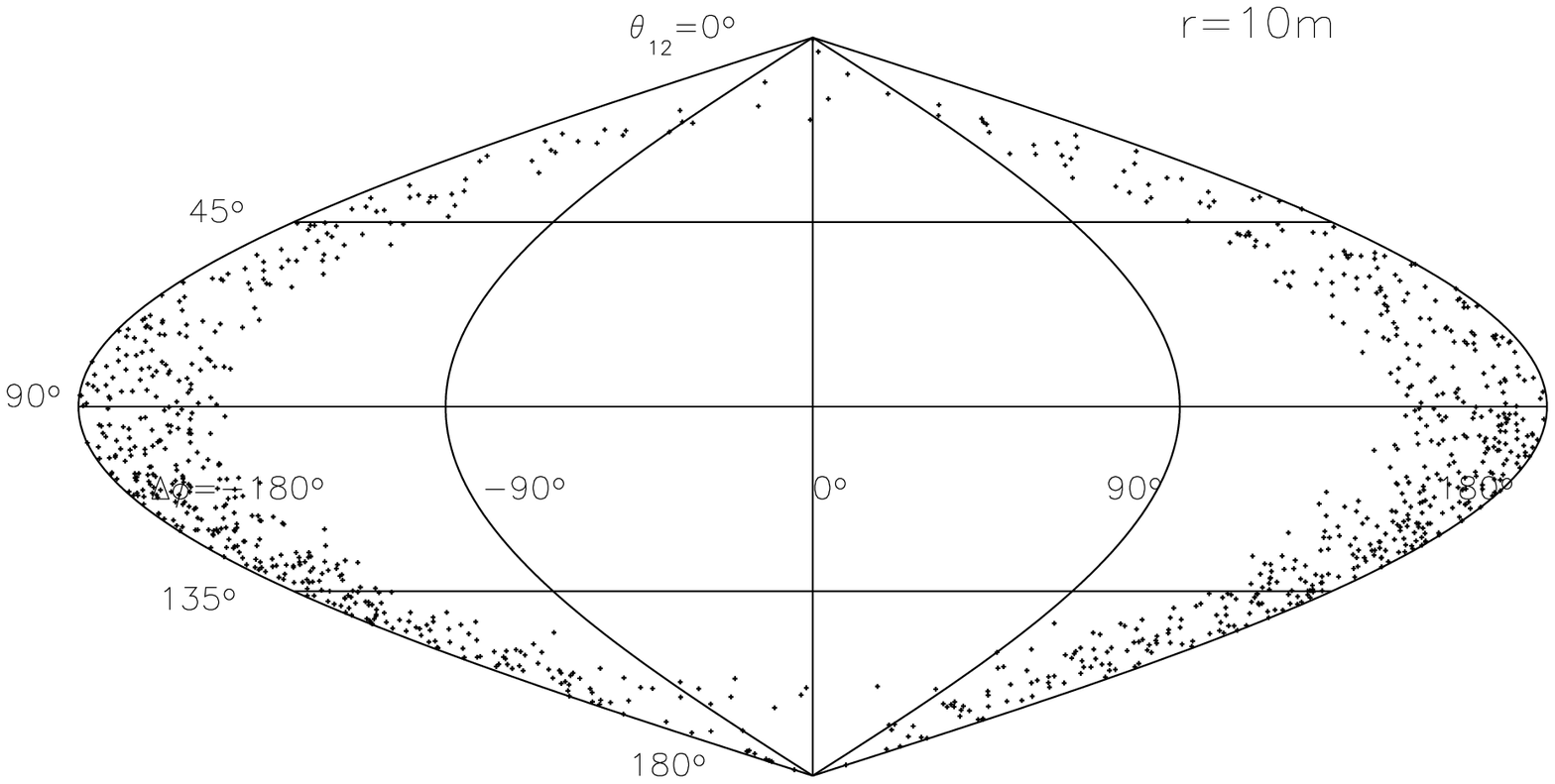}}
\end{center}
\end{figure}

The net effect of this spin evolution can be seen clearly in Figures
\ref{sinalgn} and \ref{sinanti}. We show sinusoidal projections of the
spin coordinates $(\theta_{12},\Delta\phi)$ for an initially random,
uniform distribution of spin vectors $\mathbf{S}_2$. The polar angle
$\theta_{12}$ is
defined as the angle between the two spins: $\theta_{12}\equiv
\cos^{-1}(\hat{\mathbf{S}}_1\cdot\hat{\mathbf{S}}_2)$ and the
azimuthal angle $\Delta\phi$ is defined as above. A sinusoidal
projection maps a point with spherical coordinates
$(\theta,\phi)$ onto a plane with x-y coordinates given by
$(\phi\sin\theta,\theta)$ so that equal areas in the plane correspond
to equal solid angles on the sphere. This allows an accurate way of
estimating the density of the distribution function in parameter
space.

In Figure \ref{sinalgn} we take a sample of initial conditions with
$r/m=1000$, $\theta_1=10^\circ$, and masses
$m_1=0.55$, $m_2=0.45$. The relative spin orientation
$(\theta_{12},\Delta\phi)$ for the smaller black hole has a uniform
random distribution, as shown in the upper-left frame of Figure 
\ref{sinalgn}. As the binaries evolve under radiation reaction, more
of the initially unlocked spins become locked and then librate around
$\Delta\phi = 0$, as seen in the second frame of Figure \ref{sinalgn},
corresponding to $r=250m$. Once locked into an equilibrium spin
orientation, the systems move along the curves shown in Figure
\ref{equ_align}, approaching the line $\theta_2=\theta_1$ as $L_N\to
0$. This evolution is clearly evident in the last frame of Figure
\ref{sinalgn} as $\theta_{12},\Delta\phi \to 0$ and the two spin
vectors become closely aligned with each other (although not
necessarily aligned
with the orbital angular momentum). This result may have a important
impact on the question of initial conditions for compact binaries
entering the plunge and subsequent merger phase. While the general problem
of two spinning black holes merging is still an open question in
numerical relativity, it would certainly be a significant
simplification if we can assume that the spins are initially aligned.

If we take the initial orientation of the more massive black hole spin
to be retrograde with respect to the orbital angular momentum
$(\theta_1=170^\circ)$
and again evolve a uniform distribution of $\mathbf{S}_2$
orientations, we find a qualitatively different behavior. Instead of
getting locked into the $\Delta\phi=0$ equilibrium solutions of Figure
\ref{equ_align} and evolving towards $\theta_{12}=0$, the binaries
approach the $\Delta\phi = 180^\circ$ solutions of Figure
\ref{equ_anti}, where there is no strong correlation between $\theta_1$
and $\theta_2$ at late times. Thus as $L_N \to 0$ and $\Delta\phi \to
180^\circ$, the distribution in $\theta_{12}$ is roughly uniform, with
a trend towards anti-alignment with $\theta_{12}>90^\circ$. This is
seen in the evolution of the spin distribution shown in Figure
\ref{sinanti}. Similar to Figure \ref{sinalgn}, a sample of spin
orientations $(\theta_{12},\Delta\phi)$ with uniform distribution is
taken for the set of initial conditions with $r/m=1000$ and
$\theta_1 = 170^\circ$, plotted in a sinusoidal projection in the
upper-left panel. As the systems evolve under radiation reaction, they
approach an anti-aligned spin configuration with $\Delta\phi \to \pm
180^\circ$ and $\theta_{12} \approx 90^\circ-180^\circ$.

\begin{figure}[tp]
\caption{\label{prob_algn} Probability distribution for spin orientations
(a) $\cos\theta_{12}$ and (b) $\Delta\phi$ near the end of inspiral
$(r/m=10)$. The distributions are plotted for different
values of $\theta_1(t_0)$ with $r(0)=1000m$ in all cases. Both black holes
are maximally spinning with similar masses
$m_1=0.55,m_2=0.45$. When the larger spin is initially aligned with the orbital
angular momentum $(\theta_1 < 90^\circ)$, the final spins tend to
align $(\Delta \phi \to 0^\circ,\cos\theta_{12}\to 1)$.}
\begin{center}
\scalebox{0.5}{\includegraphics{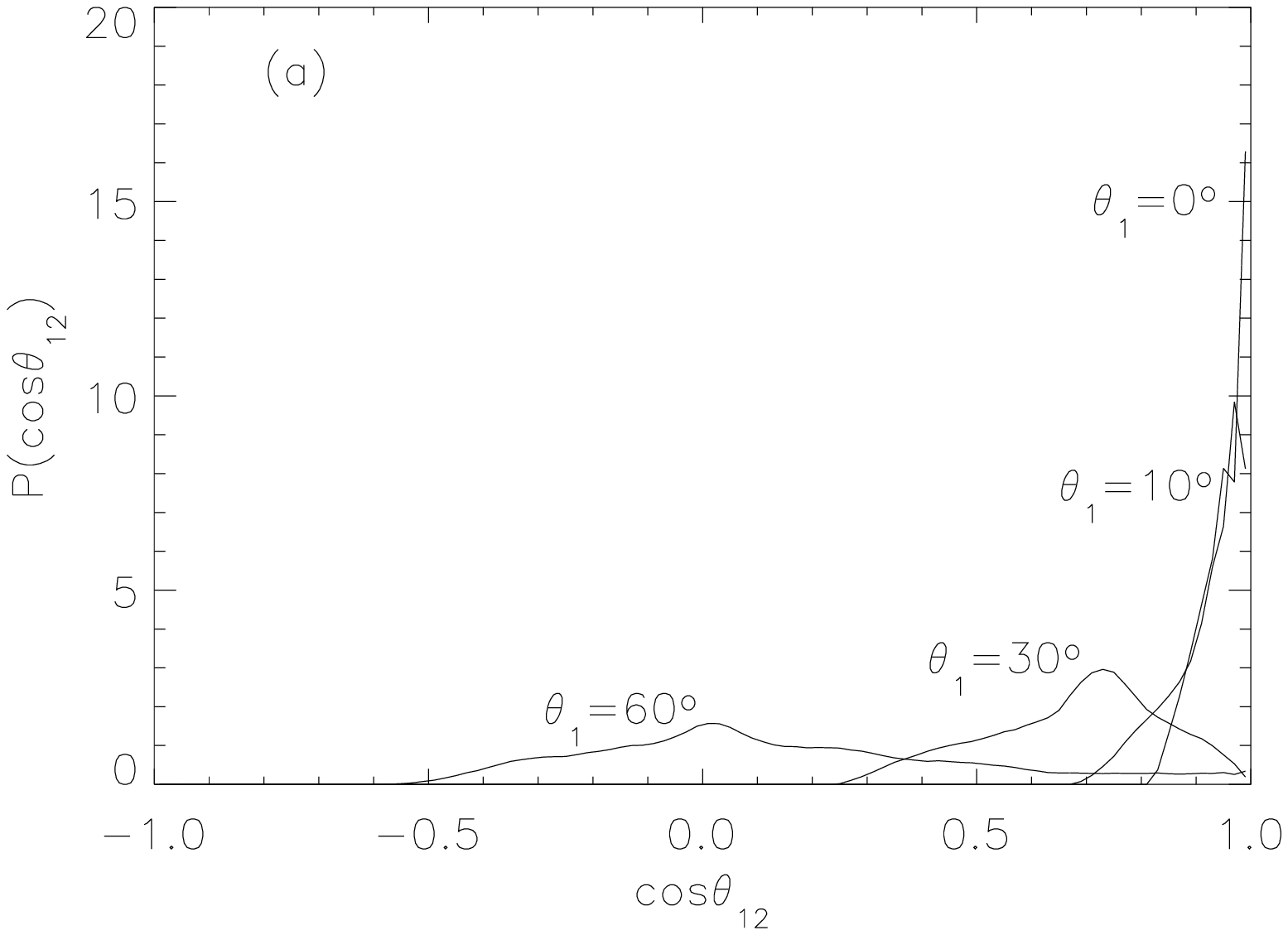}}
\scalebox{0.5}{\includegraphics{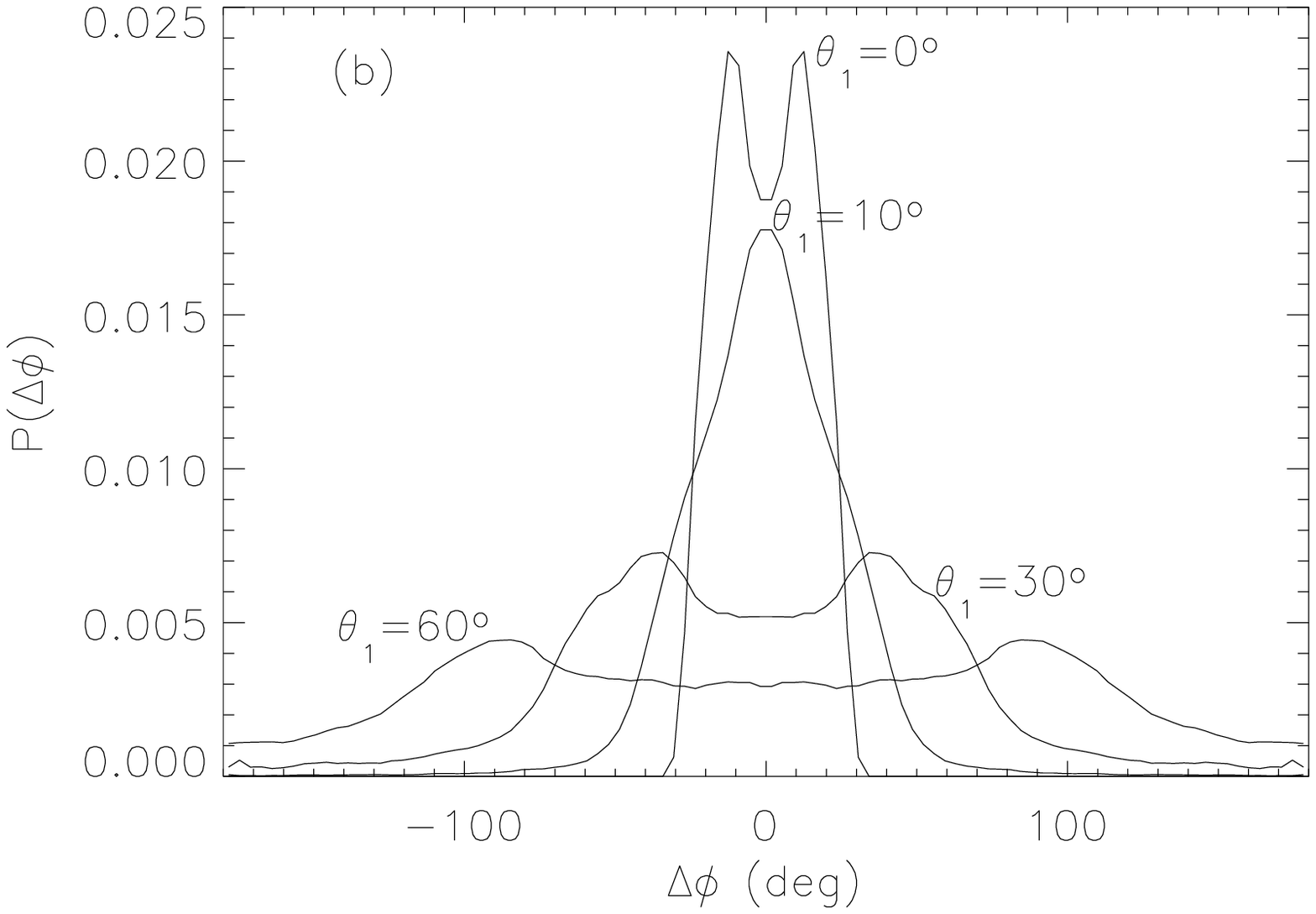}}
\end{center}
\end{figure}

\begin{figure}[tp]
\caption{\label{prob_anti} Probability distribution for spin orientations
(a) $\cos\theta_{12}$ and (b) $\Delta\phi$ near the end of
inspiral. All parameters are as in Figure \ref{prob_algn}, yet with
the larger spin initially misaligned with the orbital angular
momentum $[\theta_1(t_0) > 90^\circ]$. In these cases, the final spins also
tend to be anti-aligned with each other $(\Delta \phi \to \pm
180^\circ,\cos\theta_{12}
\to -1)$, although $\cos\theta_{12}$ is less sensitive to the initial
conditions, as can be seen from Figure \ref{sinanti}.}
\begin{center}
\scalebox{0.5}{\includegraphics{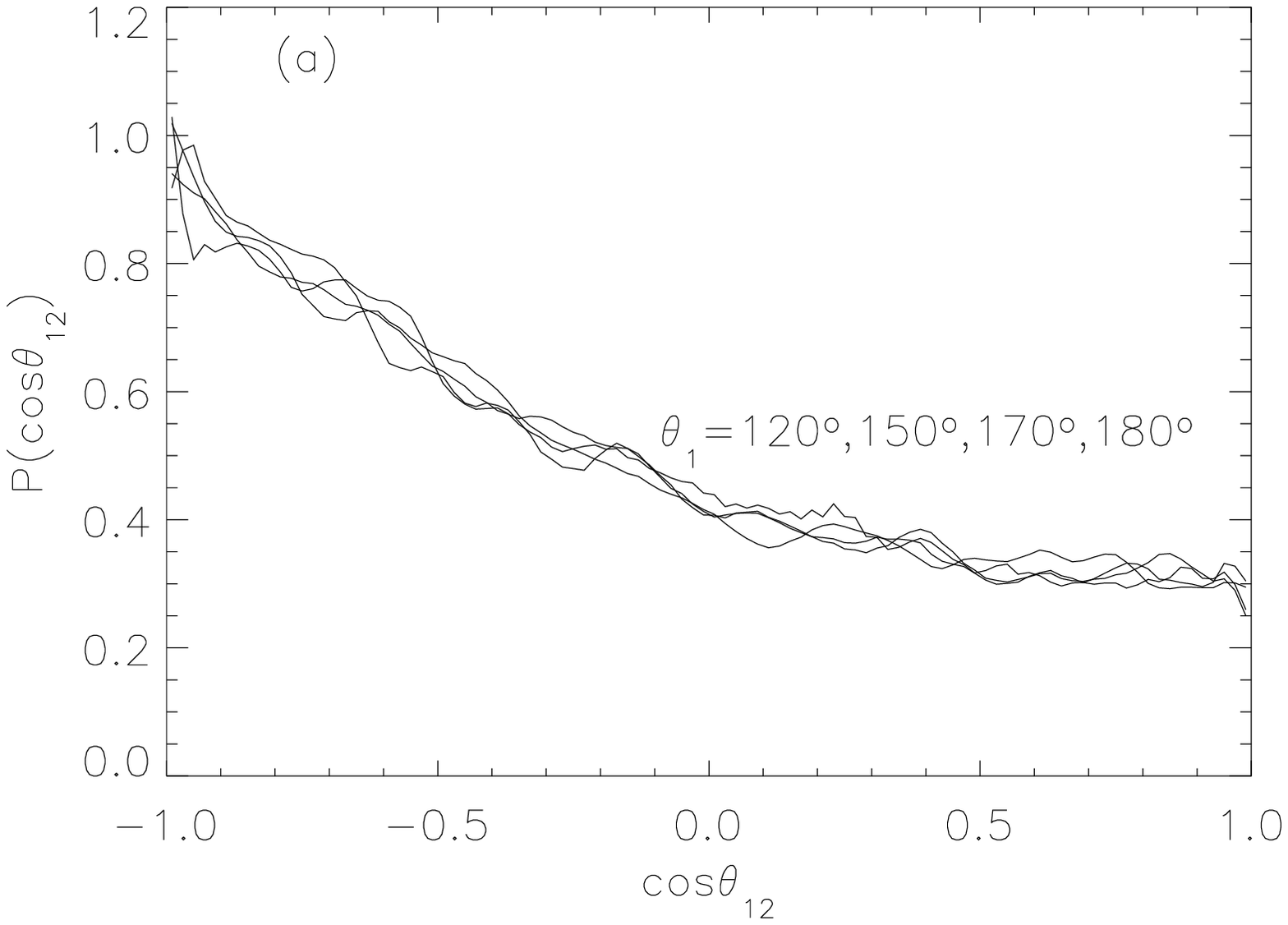}}
\scalebox{0.5}{\includegraphics{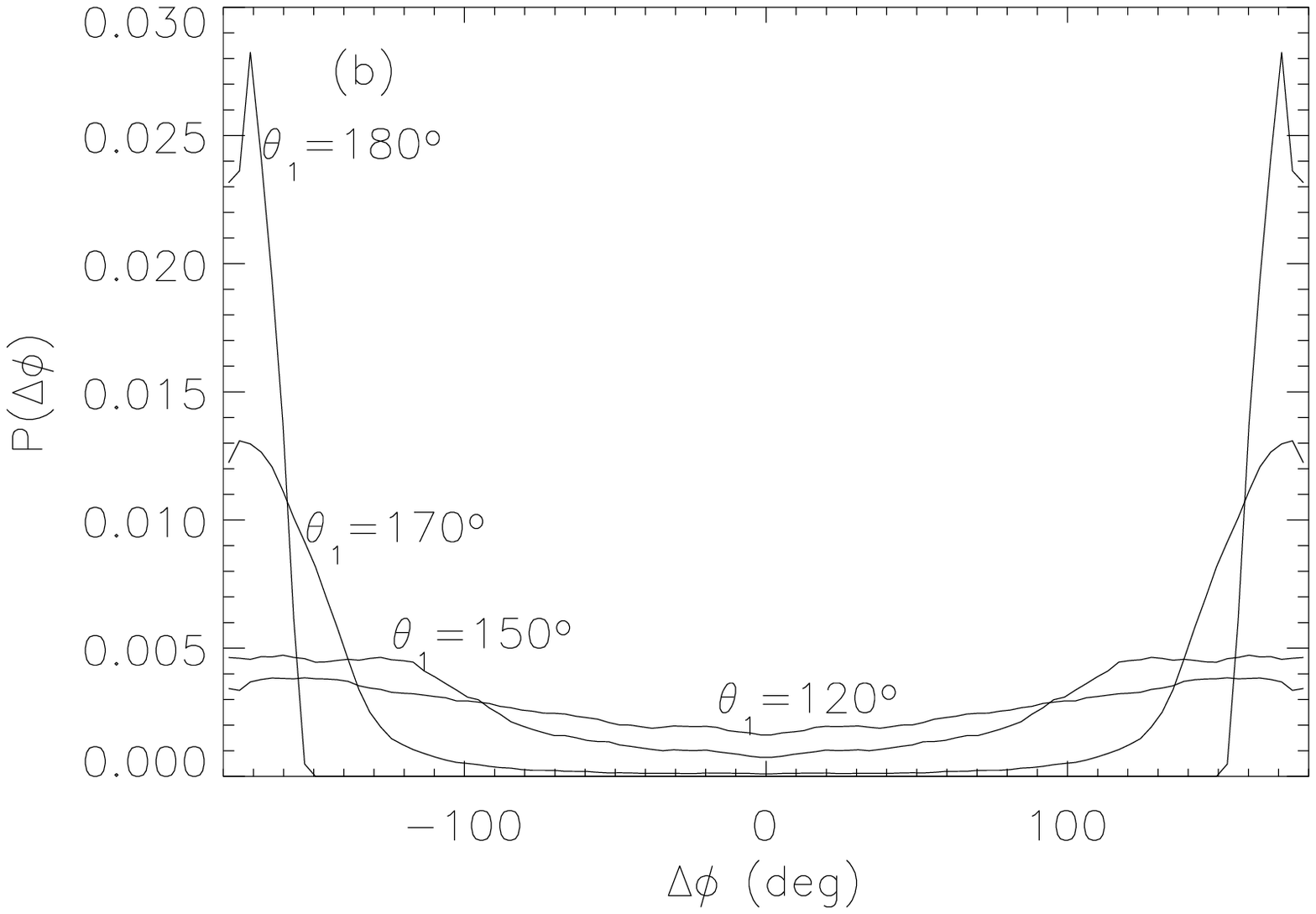}}
\end{center}
\end{figure}

We see from Figures \ref{sinalgn} and \ref{sinanti} that the final
spin configuration is strongly dependent on the initial orientation of
the larger spin with respect to the orbital angular momentum. Figures
\ref{prob_algn} and \ref{prob_anti} serve to quantify these effects
for different initial values of $\theta_1$. When $\theta_1(t_0)\approx
0^\circ$, the final spins tend to be parallel, regardless of the
initial orientation of $\mathbf{S}_2$, as shown in Figure
\ref{prob_algn}. However, if the smaller black hole initially has a
retrograde orbit, the final spins tend towards the anti-aligned
orientation described above (Figure \ref{prob_anti}). For the
intermediate cases with $\theta_1(t_0) \approx 90^\circ$, there seems
to be little net evolution in the relative spin distributions. 

Given an initial probability distribution function $f(\theta_1)$, we can
predict the final distribution of $(\theta_{12},\Delta\phi)$ by
weighting the ensemble of distributions in Figures \ref{prob_algn} and
\ref{prob_anti} by $f(\theta_1)$. This initial spin distribution
function has been the focus of much recent work in compact binary
systems \citep{kalog00,wille04}. Much of this work is based on
the evolution of two large main-sequence stars that both eventually
form compact objects through supernovae \citep{fryer99}. The resulting spin and
angular momentum vectors are largely determined by the random kick
given to each star during the asymmetric supernova explosions. Many of
these kicks will disrupt the system entirely, while the undisrupted
systems may end up with very different orientations than they had
before the supernova.  Using these binary evolution codes to give the
initial angular momentum and spin orientations, we can then use the
above results to evolve the compact binary system under radiation
reaction into the detector frequency regime. According to
\citet{kalog00}, a supernova kick velocity of $v_k \sim
200$ km/s should favorably produce a system with $\theta_1 <
30^\circ$. In this case, it follows that the $\Delta \phi =
0,\theta_{12}=0$ equilibrium solution would well describe the orientation
as the system enters the detector's frequency range. However, even for
systems locked in equilibrium, the \textit{total} spin vector may not
be parallel to the angular momentum vector, and thus simple precession
\citet{apost94} of the orbital plane will still take place during the
inspiral and must still be accounted for in the template libraries.

\section{LARGE MASS RATIOS}\label{mass_ratios}

Preliminary results with large mass ratios $(m_1:m_2 > 1000)$
show a similar behavior as the near-equal mass systems, although only
for cases where $\theta_1(t_0)$ is close to zero [or alternatively
$\theta_1(t_0)\approx 180^\circ$]. This might correspond
to a supermassive black hole in the center of a galaxy surrounded by
a large ``accretion disk'' of stellar-mass black holes and neutron
stars \citep{levin03}. If the
smaller black holes begin with randomly oriented spins, then as they evolve
towards merger, we find that $\theta_{12}$ remains roughly constant
but $\Delta\phi \to 0$, as in the previous section. However, it is
more likely that any massive stars formed in a self-gravitating
accretion disk would \textit{not} have random spin orientations, but
rather be aligned with the angular momentum of the
surrounding gas, and thus the central supermassive black hole,
simplifying the problem even further.

If, on the other hand, the stellar-mass black holes originate from an
isotropic population in the galactic bulge and sink towards the
central black hole via dynamic friction, they should have initial
velocities (i.\ e.\ orbital angular momentum) and spins with
uncorrelated, uniform random
distributions. In this case, the alignment effects described above are
unlikely to be important, and the final spin and angular momentum
orientations will likely have a similarly random distribution at the
time of merger. While the large mass ratio limits the effect of spin
locking, at the same time the spin ratio scales as $m_2^2/m_1^2$, so
to a large degree the smaller black hole can be treated as a point
mass moving in a stationary spacetime, greatly reducing the parameter search
space in any case \citep{buona03}.

For the moderate mass ratios $(m_1:m_2 \approx 1-10)$ for which
resonance behavior \textit{will} be important, one can estimate roughly
the range over which the spin locking occurs during the binary
inspiral by examining equation (\ref{s1dots2}). To leading order, we
can write the equilibrium condition as
\begin{equation}\label{approx_equ}
\frac{d}{dt}(\cos\theta_{12}) \propto
\left(\frac{m_2}{m_1}-\frac{m_1}{m_2}+
\frac{S_1\cos\theta_1-S_2\cos\theta_2}{L_N}\right)\approx 0.
\end{equation}
Taking maximal spins and $L_N = \mu m^{1/2} r^{1/2}$, we can solve for
the separation at which the resonance behavior begins to dominate the
dynamics [maximizing $r$ over the solution space to
(\ref{approx_equ})]:
\begin{equation}\label{max_r}
r_{\rm lock}/m \approx
\left(\frac{S_1\cos\theta_1-S_2\cos\theta_2}{m_1^2-m_2^2}\right)^2
\approx \left(\frac{m_1^2+m_2^2}{m_1^2-m_2^2}\right)^2.
\end{equation}
Thus we see that for large mass ratios, the spin locking does
not occur until very late in the inspiral, while for equal mass
systems it is important even for a relatively wide
separation. This result is consistent with the recent findings by
\citet{buona04} that find the best fitting factors in the cases of
nearly equal masses, further emphasizing the similarity between their
effective spin method and the equilibrium solutions of this paper.

\begin{figure}[tp]
\caption{\label{prob_mass} Probability distribution for spin orientations
(a) $\cos\theta_{12}$ and (b) $\Delta\phi$ near the end of
inspiral. The initial spin orientation has $\theta_1=10^\circ$ and
uniform distribution of $\theta_2$. Each system has maximal spins but
a different mass ratio $m_1:m_2$. The spin-locking resonance is
clearly more important for nearly equal mass systems, as seen from
equation (\ref{max_r}).}
\begin{center}
\scalebox{0.5}{\includegraphics{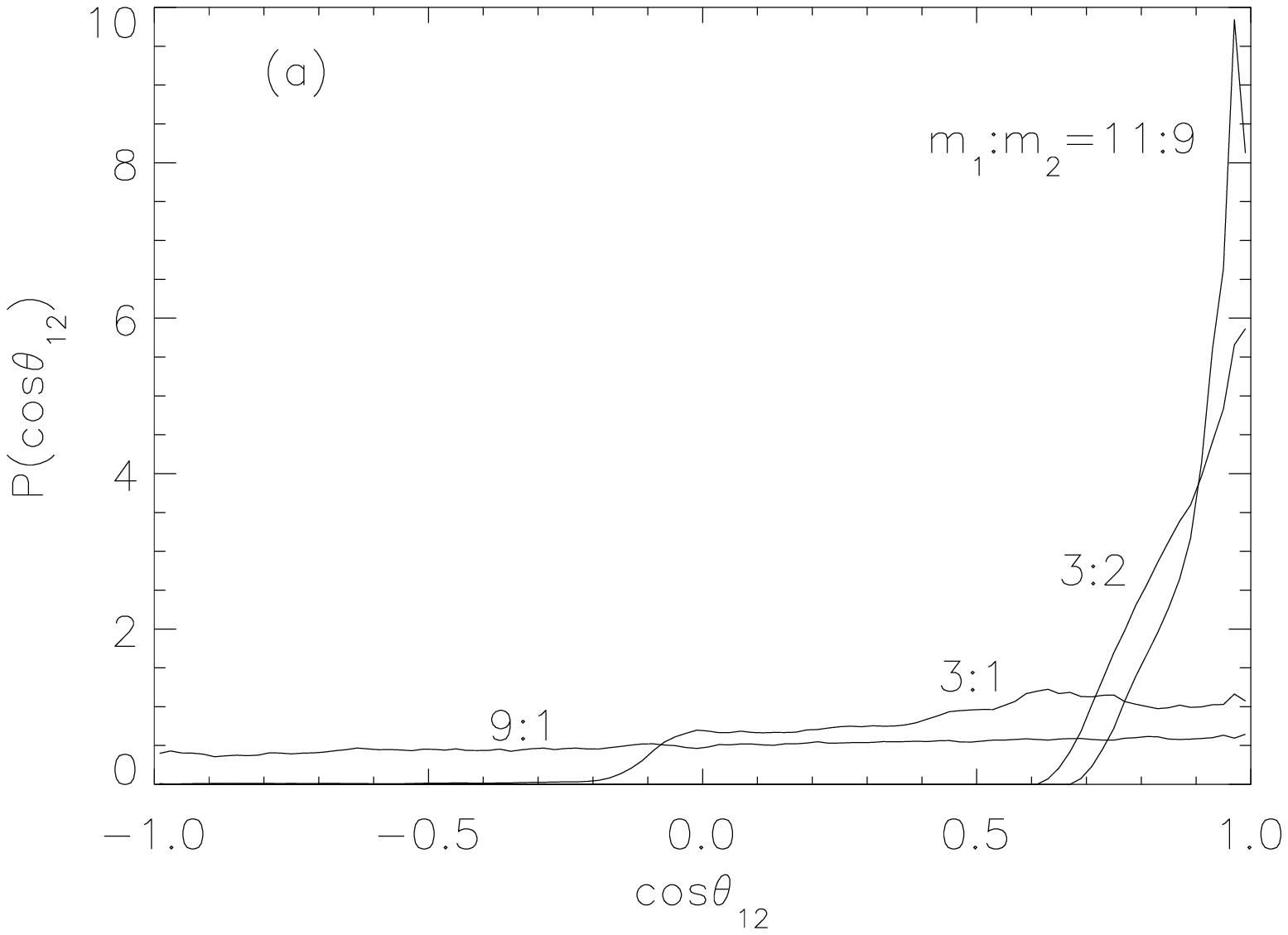}}
\scalebox{0.5}{\includegraphics{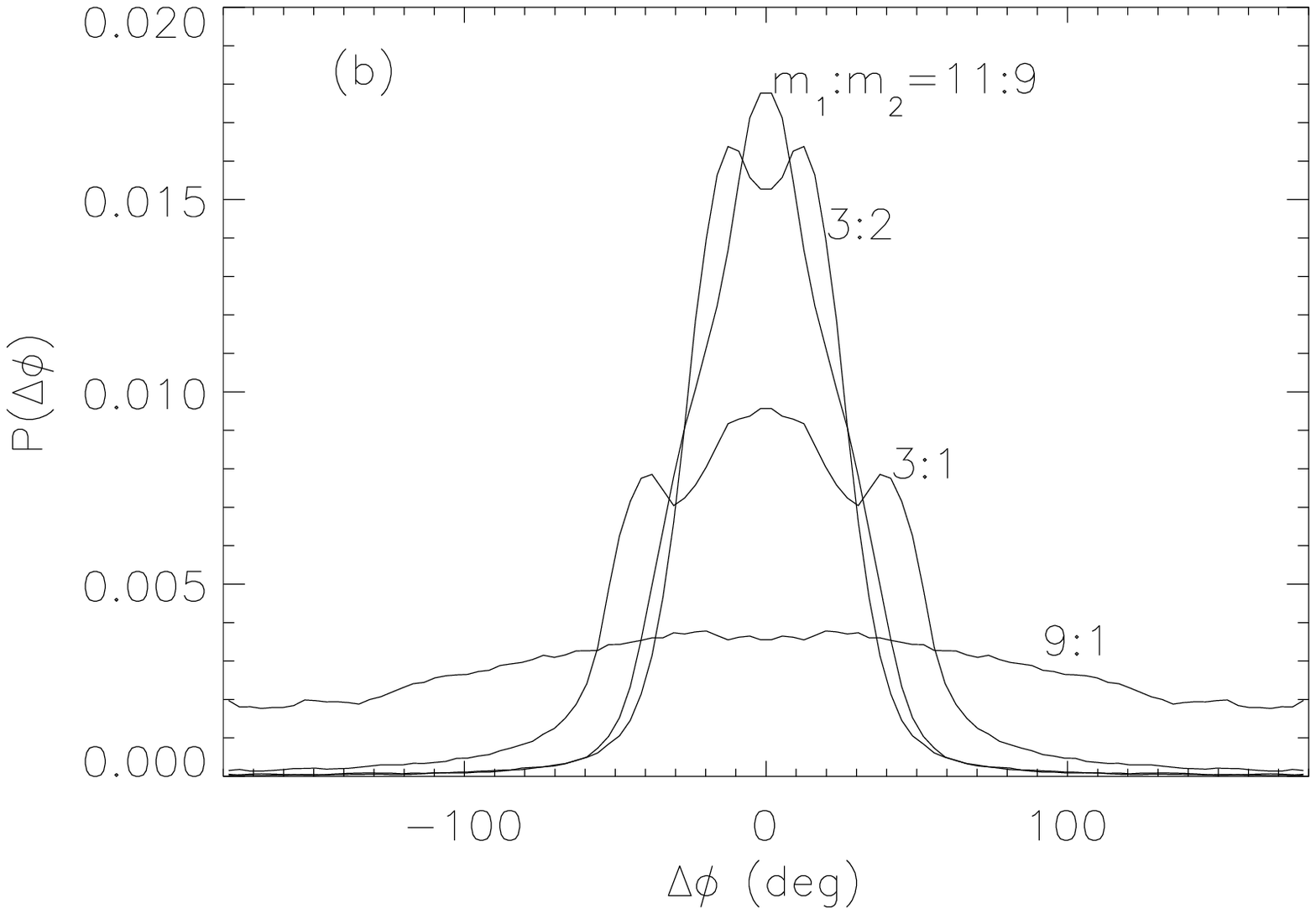}}
\end{center}
\end{figure}

The dependence of spin locking on the mass ratio is shown in Figure
\ref{prob_mass} for maximally spinning
black holes with $\theta_1(t_0)=10^\circ$. For $m_1:m_2=0.55:0.45$, the
spin locking is very strong at the end of the inspiral $(r/m=10)$, as
we saw in Figure \ref{sinalgn}. With
increasing mass ratios of $m_1:m_2=3:2$, $3:1$, and $9:1$, the resonance
behavior grows increasingly weaker, showing a smaller influence on the
initially uniform distribution of $(\theta_{12},\Delta\phi)$. However, it
should be noted that since the inspiral time scales as $\mu^{-1}$, the
higher mass ratio binaries also spend a longer time at smaller separations,
producing many more cycles in the waveform at late times. Thus the
equilibrium solutions may still give good estimates for waveform
templates even for the moderate mass ratios that would be expected for
NS-BH binaries. 

\section{EFFECTS OF ECCENTRICITY AND SPIN MAGNITUDE}\label{ecc_effects}

For the sake of simplicity, most of the computations carried out so
far have had circular orbits throughout the inspiral. However, it is
relatively straightforward to include a non-zero eccentricity and
evolve it along with the other system variables according to equations
(\ref{da_dt}, \ref{de_dt}, \ref{dL_dt}). As shown in \citet{peter64},
a high eccentricity has the effect of speeding up the inspiral time
scale. Yet as we see from equations (\ref{omg1avg}, \ref{omg2avg}),
eccentricity also increases the precession rate, accelerating the
rate of spin evolution. 

For the nearly equal mass cases discussed in Section \ref{spin_evol},
we find that
including a high initial eccentricity $(a_0=2000m, e_0= 0.9)$ reduces the
overall effect of the spin locking, but only slightly. A sinusoidal
plot of $(\theta_{12},\Delta\phi)$ analogous to Figure \ref{sinalgn}
shows the same qualitative behavior for eccentric orbits, with
marginally more scatter in the final distribution. This suggests
that the increase in the inspiral rate slightly prevails over the increased
precession rate and the spins do not have as much time to get locked
into their equilibrium states. 

While we have not yet discovered a BH-BH binary, the known
NS-NS systems show a range of eccentricities up to $e\approx 0.6$,
presumably due to the variation of initial kick velocities
\citep{wille04}. By integrating (\ref{da_dt}) and (\ref{de_dt}), it is
clear that all these known systems will circularize long before the spin
locking becomes important, and certainly before entering the detector
band \citep{belcz02}.

There is also the possibility of forming high eccentricity BH-BH
binaries through many-body exchange interactions in the cores of dense
globular clusters, which are then ejected and merge within a Hubble
time \citep{porte00}. Yet even for these systems, the merger time
scale is long enough that the circular orbit approximation is most
likely a reasonable one. In the case of the high mass ratios
described in Section \ref{mass_ratios}, high eccentricities are much
more likely. However, as was mentioned above, in such a system the
contribution of the smaller spin is
negligible and the solar mass black hole can be treated accurately as
a point mass. Furthermore, from (\ref{max_r}) we see that spin effects
are important at a much later stage for high mass ratios, giving more
time for the system to circularize, again justifying the circular
orbit approximation.

\begin{figure}[tp]
\caption{\label{prob_spin} Probability distribution for spin orientations
(a) $\cos\theta_{12}$ and (b) $\Delta\phi$ near the end of
inspiral. The initial spin orientation has $\theta_1=10^\circ$ and
uniform distribution of $\theta_2$. Each system has
$m_1=0.55,m_2=0.45$ but different values for the black hole spin
parameter $a/m$. The evolution is largely independent of the spin
magnitude for moderately values of $a/m >0.5$, while even systems with
small spin parameters exhibit significant spin locking.}
\begin{center}
\scalebox{0.5}{\includegraphics{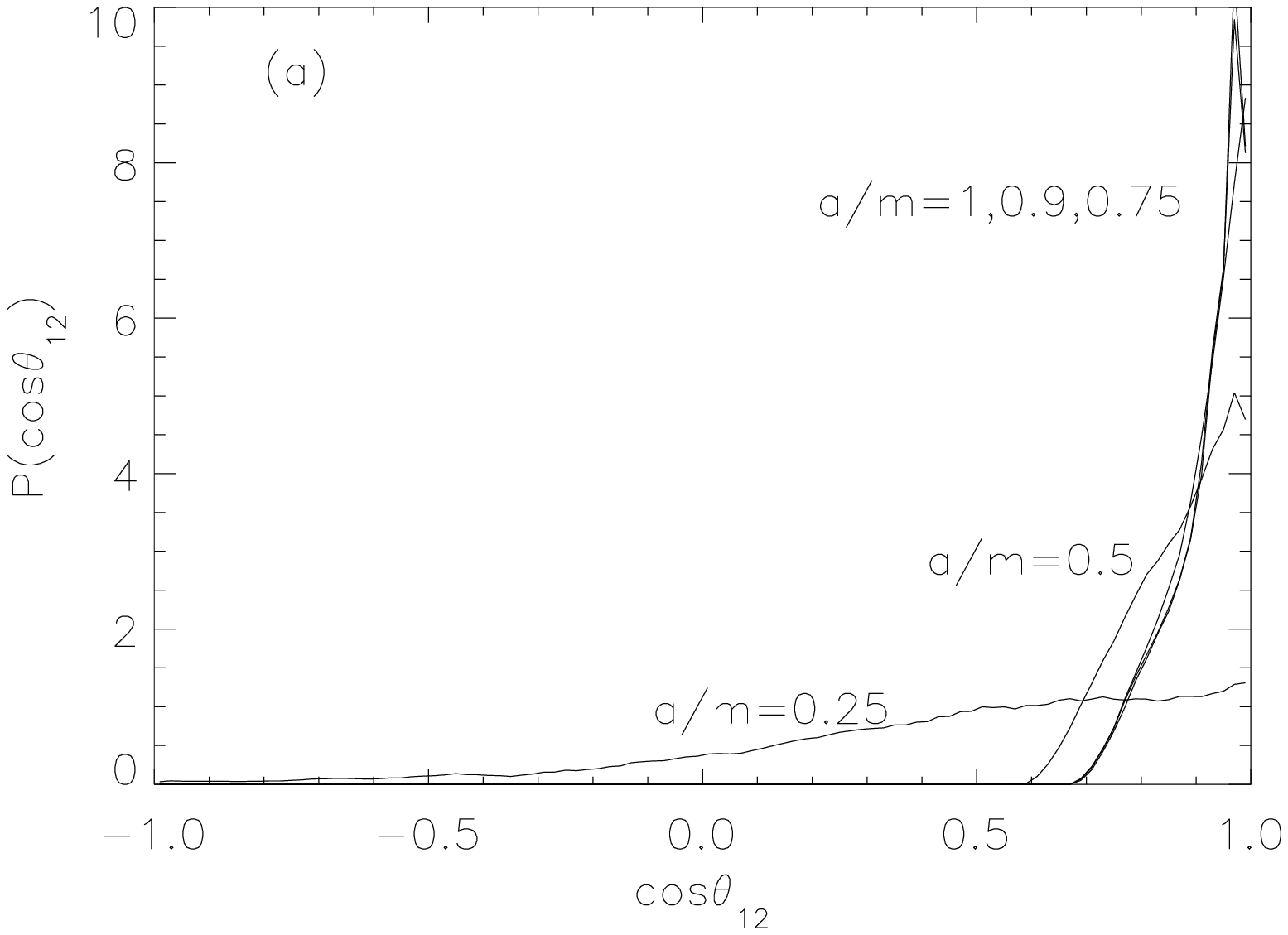}}
\scalebox{0.5}{\includegraphics{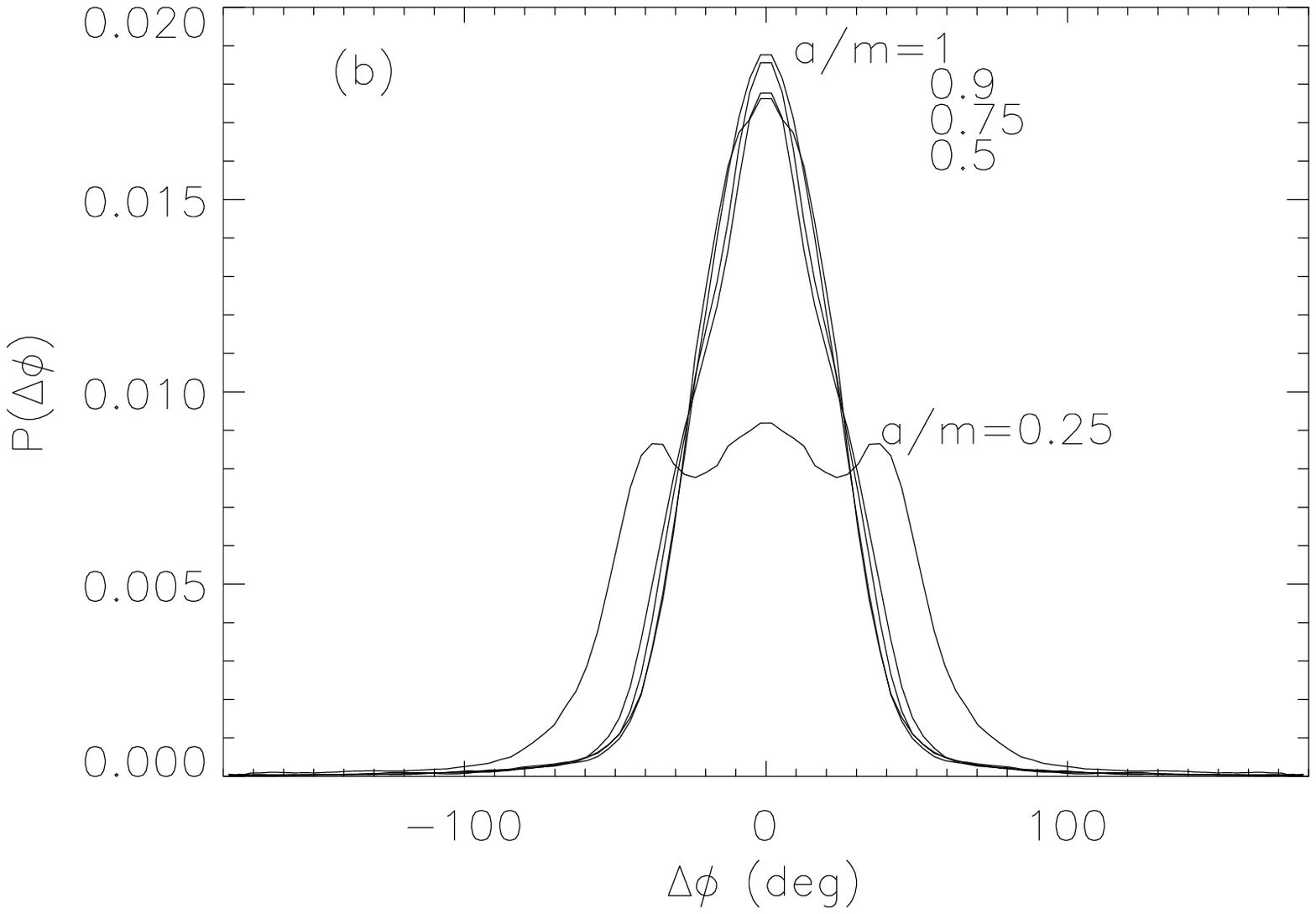}}
\end{center}
\end{figure}

We have also investigated the effects of non-maximal spin parameters
on the evolution of the spin distributions. While equation
(\ref{approx_equ}) suggests that the resonant locking effects might
fall off for smaller values of $S_1$ and $S_2$, in practice we find a
relatively weak dependence on the magnitude of the spin parameter for
$a/m=S/m^2>0.5$. Figure \ref{prob_spin} shows the probability
distributions of $(\theta_{12},\Delta\phi)$ near the end of inspiral for
a variety of spin magnitudes. Each system has the same mass ratio
$m_1:m_2=11:9$ and initial spin-orbit angle
$\theta_1(t_0)=10^\circ$. Even for spins as small as $a/m=0.25$ we see
significant spin locking at the end of inspiral. Interestingly, the
final distribution for $a/m=0.5$ and $m_1:m_2=11:9$ is almost
identical to that of $a/m=1$ and $m_1:m_2=3:1$ (cf.\ Figs.\
\ref{prob_mass} and \ref{prob_spin}), perhaps pointing to another
relation in the evolution equations that could be used to reduce
further the dimensionality of the total search space, analogous to the
effective spins described in \citep{buona04}.

\section{CONCLUSIONS}
We have derived a set of orbit-averaged post-Newtonian equations of
motion for two spinning point masses throughout the binary
inspiral. This makes it possible to model the evolution of the
relative spin orientations over a
large range of time scales. One important result is the discovery of
nontrivial equilibrium solutions that allow the system to remain
locked in a given orientation, exhibiting stable resonance
behavior. Through the loss of energy and angular momentum to
gravitational radiation, a binary system can undergo a transition from
an unlocked to locked position, in turn significantly affecting the
distribution function of the spins in parameter space.

This behavior appears to be due in large part to the constraints
placed on the orbital angular momentum and spin vectors within the
precession framework. Conservation of total angular momentum
$\mathbf{J}=\mathbf{L}+\mathbf{S}$ restrict the available regions of parameter
space accessible to the system over short time scales. In this
context, it may be more helpful to think of the spin interactions as
``spin-orbit-spin'' coupling. The precession of $\mathbf{S}_1$ will change
the orientation of $\mathbf{L}$, which in turn affects the precession
of $\mathbf{S}_2$. In an effort to better understand the
origin of this behavior, we have repeated many of the evolution
calculations, now with the direct spin-spin precession terms (order
1.5PN) turned off. This can
be done trivially by removing the terms proportional to spin in
equations (\ref{omg1avg},\ref{omg2avg}). Interestingly, the effect is
almost negligible, giving qualitatively the same results shown in
Figures \ref{sinalgn} and \ref{sinanti}. However, we should be careful
to note that this is not quite the same condition as applied in
\citep{buona03,pan04}, where the analysis is limited to spin-orbit
interactions simply by disregarding the spin of the smaller body. The
approach here is more closely analogous to the effective spin method
of \citet{buona04}. We believe their high level of success in using a
single pseudo-physical spin may be closely related to the resonances
described in this paper. 

Coupled with an estimate for the initial spin-orbit orientation from
binary evolution codes, the analysis presented in this paper can be
used to predict the distribution of spins relevant for gravitational
wave detectors and waveform templates. By limiting the template
library to the family of equilibrium solutions, we could greatly
reduce the size of the parameter search space, in turn increasing the
chances of signal detection. The spin orientation at
small orbital separations, shortly before merger, could also be very
useful information for studying the highly relativistic plunge regime,
where the post-Newtonian approximation breaks down. Among other
difficulties, the problem of black holes merging in numerical
relativity is plagued by a lack of insight into the appropriate
initial conditions. The results of this paper may serve to provide
some of that insight.

Directions for future work include modeling a vastly expanded
parameter space and in turn studying the resonance behavior over a
greater region of astrophysical interest. These studies should include
both LIGO and LISA type sources, as well as continuing to investigate
other evidence for spin-orbit and spin-spin interactions, such as the
electromagnetic signatures of
supermassive black hole mergers and binary pulsar systems. Similarly,
comprehensive calculations of compact binary formation scenarios are
needed to provide reasonable initial data for the inspiral evolution.

As we have mentioned repeatedly throughout this paper, a major part of
the detection strategy for gravitational wave observatories is that of template
matching. We propose using the equilibrium solutions as a subset of
the parameter space as a way to maximize the computational efficiency
of matched filtering. A next step would be to quantify this
presumption by calculating the fitting factors of waveforms from
off-resonance and near-resonance systems compared to those of the
exact equilibrium solutions, similar to the analysis of
\citep{buona04}. This approach would also provide concrete
predictions for detection rates as a function of the source
distributions and could be used to give stronger upper limits on
binary coalescence rates. By using this set of physical templates,
the intrinsic binary parameters of the compact system should be
determined with greater precession than attainable by using a
psuedo-physical template family.

Aside from the potentially large impact on gravitational wave
detection, the spin locking result is also of considerable interest
for general dynamical systems exhibiting resonance behavior. We have made some
progress in understanding the geometry of this resonance and how
systems can be captured into it via radiation reaction. A more
comprehensive study of the dynamics, including the possible
identification of other resonances not yet discovered, could in turn
predict new signatures to look for in gravitational wave signals. Of
particular interest would be the possibility of finding overlapping
resonance regions, classically an important breeding ground for
chaotic systems \citep{sussm01}. One
promising approach to the resonance problem is the use of conservative
Hamiltonian equations of motion (e.\ g.\ \citep{damou88,jaran98,damou01}), as
opposed to the
Lagrangian methods used here and in most other post-Newtonian
calculations. The Hamiltonian approach more closely resembles the
classical formulations of dynamic systems like the Solar System and
coupled oscillators that show many examples of resonance behavior
\citep{murra99}, and thus could give us greater insight into the
dynamics of compact binaries.

\section{Acknowledgments}
I would like to thank Alessandra Buonanno, Scott Hughes, and Vicky
Kalogera for their very detailed and helpful comments. Many
thanks to Edmund Bertschinger for his continued support and
encouragement. This work was supported in part by NASA grant
NAG5-13306. 

\appendix
\section{Derivation of the Evolution Equations}\label{app_a}
Assuming elliptical
orbits with semimajor axis $a$ and eccentricity $e$, we define a
coordinate system with the origin at the occupied focus,
$\hat{\mathbf{e}}_z = \mathbf{L}_N/|\mathbf{L}_N|$, and
$\hat{\mathbf{e}}_x$ aligned with the ellipse's pericenter. Then
$\mathbf{r}$ is given by
\begin{equation}
\mathbf{r} = [r \cos f, r \sin f, 0],
\end{equation}
where $f$ is the true anomaly. From the standard binary relations for
elliptical orbits, we define the specific angular momentum of the
reduced two-body system as
\begin{equation}
l \equiv r^2 \dot{f}=\sqrt{ma(1-e^2)}, 
\end{equation}
with
\begin{equation}
r = \frac{a(1-e^2)}{1+e\cos f}
\end{equation}
and the total mass $m=m_1+m_2$.
We can then change variables from $dt\to df$ to get
\begin{equation}
dt = \frac{[a(1-e^2)]^{3/2}}{m^{1/2}(1+e\cos f)^2}df.
\end{equation}
Averaging the $\mathbf{r}$-dependent part of equations
(\ref{omega_1}, \ref{omega_2}) over one binary period $P$ gives 
\begin{eqnarray}
\frac{1}{P}\int_0^P \frac{dt}{r^3}(\hat{\mathbf{r}}\cdot\mathbf{S})
\hat{\mathbf{r}} &=& \frac{1}{2\pi a^{3/2}}\int_0^{2\pi}\frac{df(1+e\cos
f)}{[a(1-e^2)]^{3/2}} 
(S^x\cos f+S^y\sin f)(\cos f \hat{\mathbf{e}_x} + \sin f \hat{\mathbf{e}}_y)
\nonumber\\ 
&=& \frac{\mathbf{S}-(\hat{\mathbf{e}}_z \cdot
\mathbf{S})\hat{\mathbf{e}}_z}{2a^3(1-e^2)^{3/2}}.
\end{eqnarray}
Then the orbit-averaged precession vectors are
\begin{equation}
\mathbf{\bar{\Omega}}_1 =
\frac{1}{P}\int_0^P dt \mathbf{\Omega}_1 = \frac{1}{a^3(1-e^2)^{3/2}}
\left[\left(2+\frac{3}{2}\frac{m_2}{m_1}-\frac{3}{2}\frac{\mathbf{S}_2\cdot
\mathbf{L}_N}{L_N^2}\right)\mathbf{L}_N +\frac{1}{2}\mathbf{S}_2\right]
\end{equation}
and
\begin{equation}
\mathbf{\bar{\Omega}}_2 =
\frac{1}{P}\int_0^P dt \mathbf{\Omega}_2 = \frac{1}{a^3(1-e^2)^{3/2}}
\left[\left(2+\frac{3}{2}\frac{m_1}{m_2}-\frac{3}{2}\frac{\mathbf{S}_1\cdot
\mathbf{L}_N}{L_N^2}\right)\mathbf{L}_N +\frac{1}{2}\mathbf{S}_1\right].
\end{equation}

As explained in the text, we are primarily interested in the
\textit{relative} orientation of the three vectors $\mathbf{L}_N$,
$\mathbf{S}_1$, and $\mathbf{S_2}$, a problem that can be reduced to
only four variables $(L,\theta_1,\theta_2,\Delta\phi)$ in the case of
circular orbits. To avoid certain sign ambiguities, it is actually
more convenient to use five variables, defined as following
(normalizing to $m=1$):
\begin{subequations}
\begin{eqnarray}
L &\equiv& |\mathbf{L}_N| = \mu r^{1/2}, \\
z_1 &\equiv& \cos\theta_1 = \frac{\mathbf{S}_1\cdot\mathbf{L}_N}{S_1L}, \\
z_2 &\equiv& \cos\theta_2 = \frac{\mathbf{S}_2\cdot\mathbf{L}_N}{S_2L}, \\
\beta &\equiv& \cos\theta_{12} = 
\frac{\mathbf{S}_1\cdot\mathbf{S}_2}{S_1 S_2}, \\
\alpha &\equiv& 
\frac{\mathbf{L}_N\cdot(\mathbf{S}_1\times\mathbf{S}_2)}{L S_1 S_2}.
\end{eqnarray}
\end{subequations}
Under radiation reaction, the separation $r$ will monotonically
decrease throughout the evolution, allowing us to define a new time
variable $\tau \equiv r_0-r$ so that, from (\ref{dr_dt}),
\begin{equation}
\frac{d}{dt} = \frac{64\mu}{5r^3}\frac{d}{d\tau}.
\end{equation}
In these coordinates, with an over-dot $\dot{}$ representing
$d/d\tau$, the equations of motion are:
\begin{subequations}\label{reduced_eom}
\begin{eqnarray}
\dot{L} &=& -\frac{\mu^2}{2L}, \\
\dot{z_1} &=& \frac{15}{128\mu}\alpha S_2
  \left(\frac{1}{m_2}-\frac{S_1 z_1}{L}\right), \\
\dot{z_2} &=& \frac{15}{128\mu}\alpha S_1
  \left(\frac{S_2 z_2}{L}-\frac{1}{m_1}\right), \\
\dot{\beta} &=& \frac{15}{128\mu}\alpha L
  \left(\frac{m_2}{m_1}-\frac{m_1}{m_2}+\frac{S_1z_1 -
  S_2z_2}{L}\right), \\
\dot{\alpha} &=& \frac{1}{\alpha} \left[z_1z_2\dot{\beta} +
  \beta(z_1\dot{z}_2 + z_2\dot{z}_1) - z_1\dot{z}_1 - z_2\dot{z}_2 -
  \beta\dot{\beta}\right].
\end{eqnarray}
\end{subequations}
From these functions, the original variable $\Delta\phi$ can be easily
restored through
\begin{equation}
\Delta\phi = \tan^{-1}[\alpha, (\beta-z_1z_2)].
\end{equation}
For reference, equations (\ref{reduced_eom}b-d) can be compared with
similar results derived in \citep{apost96}, equations (2a-c) and
\citep{buona04}, equations (17-19).

\newpage

\newpage

\begin{thebibliography}{99}
\bibitem[Apostolatos(1995)]{apost95} T.\ A.\ Apostolatos, Phys.\ Rev.\ D {\bf
52}, 605 (1995).
\bibitem[Grandclement, Kalogera, \& Vecchio(2003)]{grand03a}
P.\ Grandclement, V.\ Kalogera, and A.\ Vecchio, Phys.\ Rev.\ D
{\bf 67}, 042003 (2003).
\bibitem[Buonanno, Chen, \& Vallisneri(2003)]{buona03} A.\ Buonanno,
Y.\ Chen, and M.\ Vallisneri, Phys.\ Rev.\ D {\bf 67}, 104025 (2003).
\bibitem[Grandclement \& Kalogera(2003)]{grand03b}
P.\ Grandclement and V.\ Kalogera, Phys.\ Rev.\ D {\bf 67}, 082002 (2003).
\bibitem[Grandclement et al.(2004)]{grand04} P.\ Grandclement, M.\ Ihm, V.\
Kalogera, and K.\ Belczynski, Phys.\ Rev.\ D {\bf 69}, 102002 (2004).
\bibitem[Pan et al.(2004)]{pan04} Y.\ Pan et al., Phys.\ Rev.\ D {\bf 69},
104017 (2004).
\bibitem[Buonanno et al.(2004)]{buona04} A.\ Buonanno,
Y.\ Chen, Y.\ Pan, and M.\ Vallisneri, Phys.\ Rev.\ D, submitted
(2004). [gr-qc/0405090]
\bibitem[Apostolatos et al.(1994)]{apost94} T.\ A.\ Apostolatos, C.\
Cutler, G.\ J.\ Sussman, and K.\ S.\ Thorne, Phys.\ Rev.\ D {\bf 49},
627 (1994).
\bibitem[Apostotalos(1996)]{apost96} T.\ A.\ Apostolatos, Phys.\ Rev.\
D {\bf 54}, 2438 (1996).
\bibitem[Kalogera(2000)]{kalog00} V.\ Kalogera, ApJ {\bf 541}, 319 (2000).
\bibitem[Belczynski, Kalogera, \& Bulik(2002)]{belcz02} K.\ Belczynski,
V.\ Kalogera, and T.\ Bulik, ApJ {\bf 572}, 407 (2002).
\bibitem[Pfahl et al.(2002)]{pfahl02} E.\ Pfahl et al., ApJ
{\bf 574}, 364 (2002).
\bibitem[Nutzman et al.(2004)]{nutzm04} P.\ Nutzman et al., ApJ
accepted (2004). [astro-ph/0402091]
\bibitem[Portegies-Zwart \& McMillan(2000)]{porte00} S.\ F.\
Portegies-Zwart and S.\ L.\ W.\ McMillan, ApJ {\bf 528}, L17 (2000).
\bibitem[Lyne et al.(2004)]{lyne04} A.\ G.\ Lyne et al., Science {\bf 303},
1153 (2004).
\bibitem[Willems \& Kalogera(2004)]{wille04} B.\ Willens and V.\ Kalogera,
ApJ {\bf 603}, L101 (2004).
\bibitem[Merritt \& Ekers(2002)]{merri02} D.\ Merritt and R.\ D.\ Ekers,
Science {\bf 297}, 1310 (2002).
\bibitem[Hughes \& Blandford(2003)]{hughe03} S.\ A.\ Hughes and 
R.\ D.\ Blandford, ApJ {\bf 585}, L101 (2003).
\bibitem[Kidder(1995)]{kidde95} L.\ E.\ Kidder, Phys.\ Rev.\ D
{\bf 52}, 821 (1995).
\bibitem[Baker, Campanelli, \& Lousto(2002)]{baker02} J.\ Baker,
M.\ Campanelli, and C.\ O.\ Lousto, Phys.\ Rev.\ D {\bf 65}, 044001 (2002).
\bibitem[Favata, Hughes, \& Holz(2004)]{favat04} M.\ Favata, S.\ A.\ Hughes,
and D.\ E.\ Holz, ApJ {\bf 607}, L5 (2004).
\bibitem[Kidder, Will, \& Wiseman(1993)]{kidde93} L.\ E.\ Kidder, C.\ M.\
Will, and A.\ G.\ Wiseman, Phys.\ Rev.\ D {\bf 47}, 4183 (1993).
\bibitem[Peters(1964)]{peter64} P.\ C.\ Peters, Phys.\ Rev.\ {\bf 136},
B1224 (1964).
\bibitem[Murray \& Dermott(1999)]{murra99} C.\ D.\ Murray and S.\ F.\ Dermott,
\textit{Solar System Dynamics} (Cambridge University
Press, Cambridge, 1999).
\bibitem[Sussman \& Wisdom(2001)]{sussm01} G.\ J.\ Sussman and
J.\ Wisdom, \textit{Structure and Interpretation of Classical
Mechanics} (MIT Press, Cambridge MA, 2001).
\bibitem[Damour \& Schafer(1988)]{damou88} T.\ Damour and G.\ Schafer,
Nuovo Cimento Soc.\ Ital.\ Fis.\ B {\bf 101}, 127 (1988).
\bibitem[Jaranowski \& Schafer(1998)]{jaran98} P.\ Jaranowski and
G.\ Schafer, Phys.\ Rev.\ D {\bf 57}, 7274 (1998).
\bibitem[Damour(2001)]{damou01} T.\ Damour, Phys.\ Rev.\ D {\bf 64},
124013 (2001).
\bibitem[Schnittman \& Rasio(2001)]{schni01} J.\ D.\ Schnittman and
F.\ A.\ Rasio, Phys.\ Rev.\ Lett.\ {\bf 87}, 121101 (2001).
\bibitem[Cornish \& Levin(2003)]{corni03} N.\ J.\ Cornish and J.\ Levin,
Class.\ and Quant.\ Grav.\ {\bf 20}, 1649 (2003).
\bibitem[Fryer(1999)]{fryer99} C.\ L.\ Fryer, ApJ {\bf 522}, 413 (1999).
\bibitem[Levin \& Beloborodov(2003)]{levin03} Yu.\ Levin and 
A.\ M.\ Beloborodov, ApJ {\bf 509}, L33 (2003).
\end{thebibliography}
\end{document}